\documentclass[final,3p,times]{elsarticle}
\usepackage{amssymb}
\usepackage{amsthm}
\usepackage{amsmath}
\usepackage{xfrac}
\usepackage{subcaption}

\usepackage{hyperref}
\hypersetup{colorlinks,allcolors=black}

\usepackage{cancel}
\usepackage{float}
\usepackage[figurename=Fig.,labelfont=bf]{caption}

\usepackage[ruled,vlined]{algorithm2e} %to write algorithms
\makeatletter
\newcommand*{\algotitle}[2]{%
	\stepcounter{algocf}%
	\hypertarget{algocf.title.\theHalgocf}{}%
	\NR@gettitle{#1}%
	\label{#2}%
	\addtocounter{algocf}{-1}%
}
\makeatother
\usepackage{framed}
\usepackage{multicol}
\usepackage{hyperref}
\usepackage{cleveref}

\immediate\write18{%
	makeindex -s nomencl.ist -o \jobname.nls -t \jobname.nlg \jobname.nlo%
}

\usepackage{nomencl}
\makenomenclature

\journal{arXiv}

\begin{document}
	
	\begin{frontmatter}
		
		\title{Large deformation and collapse analysis of re-entrant auxetic and hexagonal honeycomb lattice structures subjected to tension and compression}
		
		\author[add1,add2]{S. Farshbaf}
            \ead{sima.farshbaf@upc.edu}
		\author[add1,add2]{N. Dialami}
		\author[add1,add2]{M. Cervera}

		\cortext[cor1]{Corresponding authors}
		\address[add1]{Universitat Politècnica de Catalunya (UPC), Campus Norte UPC, 08034 Barcelona, Spain}
  \address[add2]{Centre Internacional de Mètodes Numèrics en Enginyeria (CIMNE), Campus Norte UPC, 08034 Barcelona, Spain}

		\begin{abstract}

               Additively manufactured auxetic structures offer desirable qualities like lightweight, good energy absorption, excellent indentation resistance, high shear stiffness and fracture toughness among others. A wide range of materials from polymers to metals can be used to fabricate these structures. In contrast to conventional materials, auxetic structures exhibit negative Poisson’s ratios. Hence, unique mechanical properties can be achieved by specific design. In this work, two types of structures, namely re-entrant auxetic and non-auxetic hexagonal honeycomb, are investigated. Large deformation analyses in both 2D plane strain and 3D are conducted using linear triangular and tetrahedral multi-field displacement-pressure elements. Hyperelastic with rate-independent plasticity constitutive models are utilized and calibrated with experimental uni-axial tensile test results. The structures are subjected to compression and tension at both transversal and longitudinal directions. The contact domain method is employed to capture both self-contact and the interaction between the structure and loading plates. The obtained results show consistency with the experimental data. The outcomes of the analyses regarding the re-entrant auxetic structure agree with the expected behavior, showing a negative value of Poisson's ratio and greater efficiency of energy absorption than the hexagonal honeycomb. By understanding the influence of the loading direction on the structural behavior, equivalent Poisson's ratio and energy absorption a reliable theoretical framework for prospective designs of the lattice materials can be established.

		\end{abstract}

		
	\end{frontmatter}
	
	%% \linenumbers
	
	%% main text

 %------------------------------------------------------------------------------
 %------------------------------------------------------------------------------

\section{Introduction}
    \label{sec1}

    Additive manufacturing (AM) has revolutionized the manufacturing industry by offering design flexibility and layer-by-layer material deposition based on digital models \cite{rivet2021experimental}. The ability of this technology to create diverse cellular solids outperforms conventional manufacturing \cite{dialami2022hybrid}.
    
    Lattice structures refer to materials or components with a periodic arrangement of interconnected struts, forming a three-dimensional network \cite{dialami2021numerical}. These structures offer high strength-to-weight ratios, efficient load distribution, and enhanced mechanical properties \cite{wu2023quasi, dialami2023computational}. In-fill structures involve filling the interior of a solid component with a lightweight lattice-like pattern, reducing weight while maintaining structural integrity. Lattice and in-fill structures have gained significant attention in recent years due to their potential for lightweight, energy absorption, and customization. 
    
    Among periodic lattice materials, additively manufactured auxetic structures have recently attracted attention due to their excellent mechanical performance. Auxetic materials, unlike traditional ones, show negative Poisson's ratios, causing them to expand transversely when stretched longitudinally and contract when compressed \cite{shirzad2022auxetic}. This characteristic depends on the structural properties and not on the raw material inherently \cite{ingrole2017design, xu2020plane, evans2000auxetic}. Lattice structures were used to fill space and connect plates and faces, resulting in lightweight and sturdy components like sandwich panels. Features such as lightweight, negative Poisson's ratios, and high energy absorption capacity \cite{zhou2024energy} make auxetic structures suitable for numerous potential applications, including medical devices \cite{shirzad2022auxetic}, protective gear and textiles \cite{krishnan2022auxetic}. Foam padding with auxetic behavior is used to offer more protection for athletes or military personnel. Besides, in order to operate advanced comfort and mobility, auxetic fabrics are used to create clothing that stretches in all directions to accommodate human growth \cite{wang2014auxetic}. The potential applications of auxetic structures are still under investigation, and many innovative products \cite{alomarah2023bio} will be produced based on their unique properties in the future.
    
    A wide range of materials can be used to fabricate auxetic structures including polymers \cite{chan1999mechanical}, metals \cite{choi1992non, friis1988negative, galati2022numerical}, ceramics \cite{dominec1992elastic}, and composites \cite{zhang2015fabrication, li2023auxetic}. Polyurethane, polystyrene and polyethylene are the commonly used materials among others to manufacture auxetic foams\cite{zhou2015superelasticity}. Metals including titanium and nickel have been employed in the fabrication of auxetic structures with high strength and durability \cite{abel2021fused}. Auxetic structures made of ceramic materials, such as zirconia, can be used in medical implants to provide better compatibility with human tissues \cite{hu2021crack}. Composites, which combine two or more materials, have also been utilized to make auxetic structures with improved properties, like increased stiffness and strength \cite{yu2023compressive}. 
    
\begin{figure}[h!]
            \centering
            \begin{subfigure}[h]{0.17\textwidth}
                \centering
                \includegraphics[width=1\textwidth]{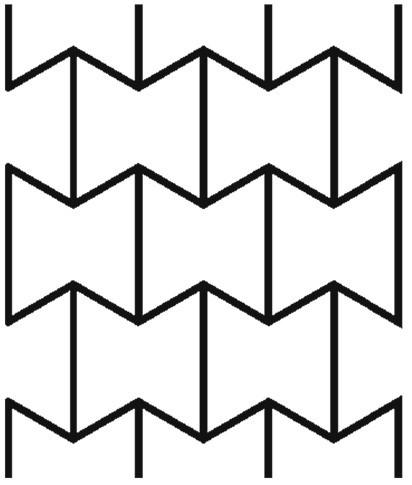}
                \vspace{0.0cm}
                \caption{Re-entrant}
                \label{fig1a}
            \end{subfigure}
            \hspace{1cm}
            \begin{subfigure}[h]{0.19\textwidth}
                \centering
                \includegraphics[width=1\textwidth]{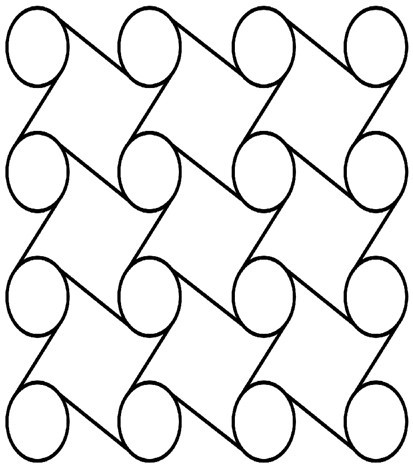}
                \caption{Chiral}
                \label{fig1b}
            \end{subfigure}
             \hspace{1cm}
            \begin{subfigure}[h]{0.22\textwidth}
                \centering
                \includegraphics[width=1\textwidth]{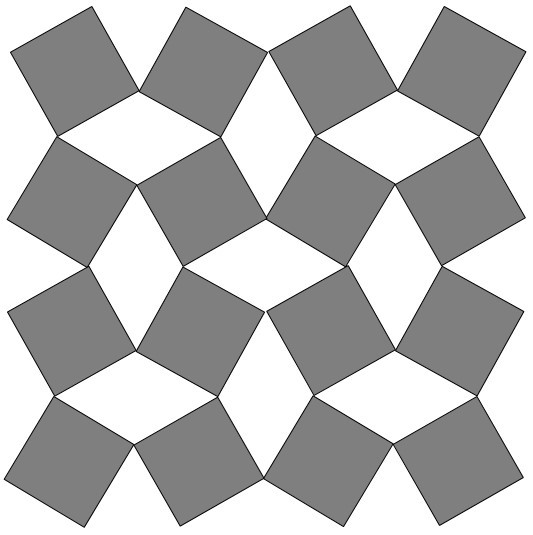}
                \caption{Rotating plates}
                \label{fig1c}
            \end{subfigure}
            \vspace{0.0cm}
            \caption{Various types of auxetic structures.}
            \label{fig1}
        \end{figure}

    Over the past few decades, various kinds of auxetic structures have been designed. The most popular and earliest structure is the re-entrant hexagonal honeycomb, designed by Gibson and Ashby in 1982 \cite{gibson1982mechanics}. Afterward, the double arrowhead configuration \cite{larsen1997design} and the STAR-systems \cite{theocaris1997negative} were designed as additional re-entrant systems. The angle of the ribs that constitute the individual cells demonstrates the auxetic properties in re-entrant structures \cite{montgomery2023elastic}. An alternative to achieve structural auxetic behavior involves considering the wrapping and unwrapping of ligaments around particular nodes, chiral structures employ this principle. Lakes \cite{lakes1991deformation} was the first to propose chiral configurations as futuristic auxetic structures. Another type of auxetic structure involves the manifestation of auxetic characteristics within the rotational degree of freedom of plates or crystals that are connected via hinges at their vertices were studied by several researchers \cite{li2023mechanical, teng2023stretchable, tao2022novel}. This positioning causes the plates to rotate when compressed or expanded in one direction, resulting in compression or expansion in the opposite direction, and eventually producing a negative Poisson’s ratio \cite{grima2000auxetic}. \autoref{fig1} depicts auxetic re-entrant, chiral and rotating plate structures. Structural dimensions of auxetic materials, such as cell size, cell shape, strut thickness, orientation, and interconnectivity extremely influence their mechanical behavior including strength, stiffness, flexibility, and elasticity \cite{montgomery2023elastic}.
    
     The orientation and interconnectivity of the cells and struts can impact the mechanical properties of the structure. Therefore, it is possible to tailor the mechanical behavior of auxetic structures in specific applications by precisely managing these structural dimensions. Auxetic structures can display assorted levels of deformation and response when subjected to loading in various directions. For example, loading along the axis of alignment of the structure’s cells or struts can result in increased stiffness and strength, while loading perpendicular to this axis can cause intensified flexibility and deformation \cite{zhou2023comparison}. The effect of loading direction depends on the structural dimensions and orientation of the auxetic structure, where careful consideration is required when designing and using these structures to ensure optimized performance in different loading conditions.
     
     The present paper aims to computationally assess the mechanical behavior of auxetic and non-auxetic structures. The impact of different quasi-static loading in different directions on the structural performance and mechanical behavior such as deformation patterns, Poisson’s ratio and energy absorption are investigated. Re-entrant Auxetic (RA) and Hexagonal Honeycomb (HH) are subjected to in-plane uni-axial tension and compression loads to analyze their performance in 2D plane strain and 3D. To accurately represent the material response, a finite strain hyperelastic with the rate-independent plastic constitutive model is used. Furthermore, friction and self-contact are taken into account to simulate the realistic behavior of compression tests. The simulation results are validated with experimental uni-axial tensile tests on polyurethane (PU) to characterize the material, and by compression tests \cite{luo2022mechanical}.

     The organization of the article is as follows. In \cref{subsec2.1}, the governing equations of the problem including momentum and mass balance are introduced and the stabilization equation for the displacement-pressure (U-P) mixed element is derived. \Cref{subsec2.2} describes the hyperelastic and plastic constitutive models used in the numerical model. Next, the contact domain method is briefly discussed in \cref{subsec2.3} with a short review of normal and tangential contact components. \Cref{subsec2.4} gives details of the finite element model setup and configuration. Afterwards, \cref{sec3} delves into the obtained results regarding deformation behavior, Poisson's ratio and energy absorption performance of the structures. Concluding remarks are presented in \cref{sec4}.

 %------------------------------------------------------------------------------
 %------------------------------------------------------------------------------

\section{Description of the numerical model}
    \label{sec2}

    \subsection{Basic equations}
    \label{subsec2.1}

         In this section, the governing equations describing the Initial Boundary Value Problem (IBVP) and the corresponding field variables based on the theory of continuum mechanics are discussed.

        \subsubsection{Balance of momentum and mass}
        \label{subsubsec2.1.1}

            The current configuration of a solid continuum is considered in which $x$ refers to the current position of a particle, $u(x,\,t)$ and $p(x,\,t)$ indicate the displacement and pressure of the particle at time $t$, respectively. The equations of the momentum and mass conservation of the solid continuum in the weak form are expressed as:

            \begin{equation}\label{eq1}
            \begin{split}
              & \int_{{{V}_{t}}}{\delta {{\varepsilon }_{ij}}{{s}_{ij}}d{{V}_{t}}+}\int_{{{V}_{t}}}{\delta {{\varepsilon }_{ij}}p{{\delta }_{ij}}d{{V}_{t}}-}\int_{{{V}_{t}}}{{{w}_{i}}{{b}_{i}}d{{V}_{t}}-}\int_{{{\Gamma }_{N}}}{{{w}_{i}}t_{i}^{p}d\Gamma =0} \\ 
             & \int_{{{V}_{t}}}{-\frac{q}{\kappa}pd{{V}_{t}}+}\int_{{{V}_{t}}}{q\frac{\ln (J)}{J}d{{V}_{t}}}=0 \\ 
            \end{split}
            \end{equation}
            where ${{s}_{ij}}$ is the deviatoric part of the Cauchy stress tensor, $p$ is the pressure, ${{b}_{i}}$ are the external body forces, $t_{i}^{p}$ are the prescribed surface forces, $\delta {{\varepsilon }_{ij}}$ refers to the virtual strain field, ${{w}_{i}}$ are the weighting functions for the displacement field and ${{V}_{t}}$ is the volume of the solid in the current configuration. Additionally, $J$, $\kappa$ and $q$ are the deformation gradient, modulus of incompressibility and weighting functions for the pressure, respectively.

            The Dirichlet and Neumann boundary conditions are:

            \begin{equation}\label{eq2}
            \begin{split}
              & \mathbf{u}(\mathbf{x},\,t)=\mathbf{\bar{u}}(\mathbf{x},\,t)\,\,\,\,\,\forall \mathbf{x}\in {{\Gamma }_{D}} \\ 
             & \mathbf{v}(\mathbf{x},\,t)=\mathbf{\bar{v}}(\mathbf{x},\,t)\,\,\,\,\,\forall \mathbf{x}\in {{\Gamma }_{D}} \\ 
             & \sigma (\mathbf{x},\,t)\cdot \mathbf{n}=\mathbf{\bar{h}}(\mathbf{x},\,t)\,\,\,\,\,\forall \mathbf{x}\in {{\Gamma }_{N}} \\ 
            \end{split}
            \end{equation}
            where $\mathbf{\bar{u}}(\mathbf{x},\,t)$, $\mathbf{\bar{v}}(\mathbf{x},\,t)$ and $\mathbf{\bar{h}}(\mathbf{x},\,t)$ denote the functions corresponding to the boundary conditions, $\mathbf{n}$ is normal in the out-ward direction of the boundary, ${{\Gamma }_{D}}$ and ${{\Gamma }_{N}}$ are Dirichlet and Neumann boundaries.

        \subsubsection{Stabilization of the U-P mixed element}
        \label{subsubsec2.1.2}

             In 2D problems, the 2-simplex element is a triangle with linear interpolations to approximate the field of main variables $u$ and $p$ for the formulation in \cref{subsubsec2.1.1}. Triangular elements with linear interpolation have some advantages including inexpensive computational cost and simple formulation. However, employing an identical order of interpolation for both the pressure and displacement fields leads to a lack of stability and spurious oscillations of the solutions. These oscillations arise due to the dissatisfaction of the Ladyzhenskaya–Babuška–Brezzi (LBB) condition for stability \cite{carbonell2020modelling}. This behavior also occurs in 3D problems for the linear tetrahedron as a 3-simplex element with linear interpolations. 

            To overcome the problem, the Polynomial Pressure Projection (PPP) stabilization method \cite{dohrmann2004stabilized} is used. The PPP incorporates a stabilization pressure term into the mass conservation equation depending on the difference between the continuous p pressure and a projection onto a discontinuous space \cite{rodriguez2018generation}. By using PPP as a stabilization technology, the mass conservation equations are re-written as:

            \begin{equation}\label{eq3}
            \int_{{{V}_{t}}}{-\frac{q}{\kappa }pd{{V}_{t}}+\frac{\ln (J)}{J}\int_{{{V}_{t}}}{q}\frac{\ln (J)}{J}d{{V}_{t}}}+\int_{{{V}_{t}}}{(q-\overset{\scriptscriptstyle\smile}{q})\frac{{{\alpha }_{s}}}{\mu }(p-\overset{\scriptscriptstyle\smile}{p})d{{V}_{t}}}=0 \\
            \end{equation}
            where ${{\alpha }_{s}}$ is the stabilization parameter (here it takes the value of 1) and $\mu$ is the shear modulus of the material. Also, $\overset{\scriptscriptstyle\smile}{p}$ and $\overset{\scriptscriptstyle\smile}{q}$ are the best approximations of the $p$ and $q$, respectively. Further information can be found in references \cite{carbonell2020modelling, rodriguez2018generation, rodriguez2016particle, rodriguez2017continuous}.

%------------------------------------------------------------------------------

    \subsection{Constitutive model}
    \label{subsec2.2}

       In this section, the hyperelastic and plastic constitutive models are described using a multiplicative decomposition of the deformation and a rate-independent hardening rule. When subjected to large deformations, auxetic structures may exhibit complex deformation patterns, which hyperelastic models can describe. In situations where auxetic structures experience extensive deformation under high loads or undergo permanent shape changes, plastic constitutive models become necessary. Therefore, in this section, an appropriate non-linear elastic with rate-independent plasticity is described. For the elastic part, the Neo-Hookean hyperelastic model and for the plastic part, the Mises-Huber yield function with Simo's exponential hardening law \cite{simo1992associative} will be used.

        \subsubsection{Hyperelasticity}
        \label{subsubsec2.2.1}

            It is widely recognized that in large deformation analysis, an effective method for defining the deformation gradient is by splitting it into hyperelastic and plastic components in a multiplicative manner \cite{simo1992associative}.
            
            The total deformation gradient, $\mathbf{F}$, is split multiplicatively into elastic and plastic parts: $\mathbf{F}={{\mathbf{F}}^{e}}{{\mathbf{F}}^{p}}$. The elastic part also can be further decomposed into its volumetric and deviatoric parts: ${{\mathbf{F}}^{e}}=\mathbf{F}_{v}^{e}\mathbf{F}_{d}^{e}$, $\mathbf{F}_{v}^{e}={{({{J}^{e}})}^{\frac{1}{3}}}\mathbf{1}$ and $\mathbf{F}_{d}^{e}={{\mathbf{F}}^{e}}{{({{J}^{e}})}^{-\frac{1}{3}}}\mathbf{1}$, respectively, where ${{J}^{e}}=\det ({{\mathbf{F}}^{e}})$ is the Jacobian of the elastic deformation gradient.

            In this context, hyperelastic deformation is considered as: 

            \begin{equation}\label{eq4}
                \begin{split}
                  & \mathbf{\tau }=2{{\rho }_{0}}{{\mathbf{b}}^{e}}{{\partial }_{{{b}^{e}}}}\psi ({{\mathbf{b}}^{e}}) \\ 
                \end{split}
            \end{equation}
            where $\mathbf{\tau }$ is the the Kirchhoff stress tensor, ${{\mathbf{b}}^{e}}={{\mathbf{F}}^{e}}{{\mathbf{F}}^{eT}}$ is the elastic left Cauchy-Green tensor and ${{\mathbf{\bar{b}}}^{e}}=\mathbf{F}_{d}^{e}\mathbf{F}_{d}^{eT}$ refers to its deviatoric part. Consequently, the elastic part of the free energy, denoted as $\psi ({{\mathbf{b}}^{e}})$, can be split into two terms: the volumetric term described by the function $U({{J}^{e}})$, and the deviatoric term denoted by $W({{\mathbf{\bar{b}}}^{e}})$, as:

            \begin{equation}\label{eq5}
                \psi ({{\mathbf{b}}^{e}})=U({{J}^{e}})+W({{\mathbf{\bar{b}}}^{e}})
            \end{equation}
            In plasticity, additional term $Z({{\varepsilon }_{p}})$ is added to the free energy equation to incorporate the plastic response of the material:

            \begin{equation}\label{eq6}
            \psi ({{\mathbf{b}}^{e}}) =U({{J}^{e}})+W({{\mathbf{\bar{b}}}^{e}})+Z({{\varepsilon }_{p}})
            \end{equation}
            The explicit forms can be regarded as the elastic-related terms of Eq. \ref{eq6} \cite{simo1992associative}:

            \begin{equation}\label{eq7}
                \begin{split}
                 & U({{J}^{e}})=\frac{1}{2}\kappa {{\ln }^{2}}({{J}^{e}}) \\ 
                 & W({{{\mathbf{\bar{b}}}}^{e}})=\frac{1}{2}\mu [tr({{{\mathbf{\bar{b}}}}^{e}})-3] 
                \end{split}
            \end{equation}
        where $\kappa >0$ and $\mu >0$ are the bulk and shear moduli, respectively.

        \subsubsection{Rate-independent plasticity}
        \label{subsubsec2.2.2}

            To computationally model the plastic behavior, a yield criterion that determines the transition of the material from the elastic to the plastic regime is introduced. The evolution of the plastic strain is described by a flow rule. A hardening law is used to characterize the evolution of the yield limit \cite{de2011computational}. Typically, in quasi-static states, only strain hardening behavior is taken into account and strain rate hardening terms are ignored. Therefore, in this work, a rate-independent plasticity model is employed where its yield function is regarded as flow potential to obtain the evolution equation. It is called associative flow rule \cite{simo1992associative} in which the direction of plastic flow aligns with the direction of the yield function.

            To represent the constitutive behavior of the auxetic structure under large strain deformations, it is important to consider an accurate flow stress model. Here, in the case of rate-independent plasticity, the classical Mises-Hubber is selected as the yield criterion. It is expressed in terms of the second invariant of the Kirchhoff stress tensor (${{J}_{2}}=\frac{1}{2}dev\tau :dev\tau$):

            \begin{equation}\label{eq8}
                \begin{split}
                  & f(\tau ,\,{{\varepsilon }_{p}})=\parallel dev(\tau )\parallel -\sqrt{\frac{2}{3}}(\sigma +\beta ) \le 0 \\ 
                \end{split}
            \end{equation}
            where $\sigma $ is the flow stress, ${\varepsilon }_{p}$ is the plastic strain,  $\beta =-{K}'({{\varepsilon }_{p}})$ denotes the isotropic nonlinear hardening modulus which depends on the hardening law used.
            
            In incremental plasticity, the flow rule is used to obtain the rate of the plastic part of the deformation gradient ${{\mathbf{F}}^{p}}$. It is in terms of the plastic flow direction tensor $\mathbf{n}=\frac{dev(\tau)}{\parallel dev(\tau) \parallel}$. In association with the yield surface, the plastic part of the deformation gradient is expressed as:

            \begin{equation}\label{eq8.1}
                {{\mathbf{F}}^{p}}=\dot{\lambda }\frac{dev(\tau)}{\parallel dev(\tau) \parallel}
            \end{equation}
            where $\dot{\lambda }$ is the plastic multiplier or the consistency parameter used in the Kuhn-Tucker conditions. The evolution of the effective plastic strain is written in terms of the plastic multiplier:

            \begin{equation}\label{eq8.2}
                {{\dot{\varepsilon }}_{p}}=\sqrt{\frac{2}{3}}\dot{\lambda }
            \end{equation}
            
\begin{figure}[h]%
            \centering
            \includegraphics[width=0.8\textwidth]{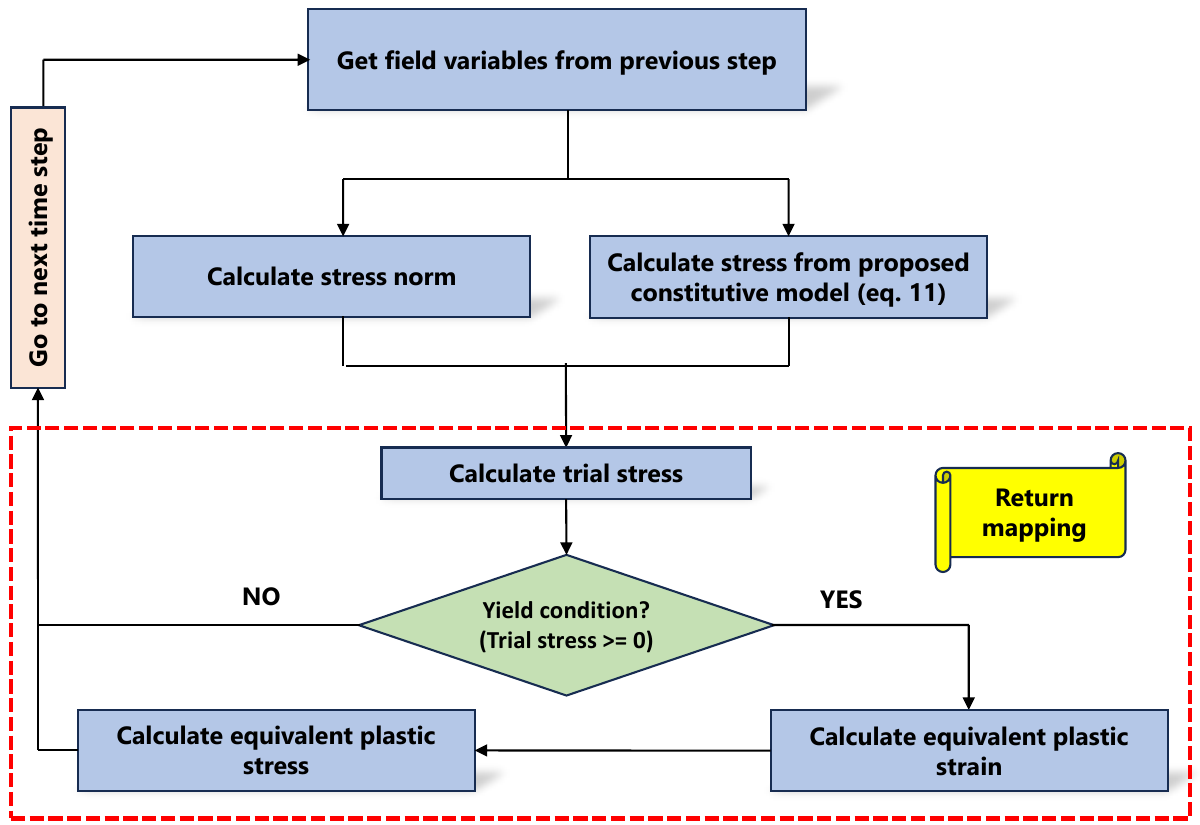}
            \caption{Flowchart of the plastic constitutive model.}\label{Constitutive_Flowchart}
            \end{figure}

            Ultimately, among numerous phenomenological and physical-based constitutive descriptions, in this study, the strain rate-independent Simo's exponential model \cite{simo1992associative} is used as the hardening law.      

            \begin{equation}\label{eq9}
            (\sigma +\beta )={{\sigma }_{0}}+K{{\varepsilon }_{p}}+({{K}_{\infty }}-{{K}_{0}})(1-{{e}^{-n{{\varepsilon }_{p}}}})
            \end{equation}
            where ${{\sigma }_{0}}$ is yield stress, $K$ is the kinematic hardening, ${{K}_{\infty }}$ indicates the infinity hardening, ${{K}_{0}}$ is reference hardening and $n$ denotes the hardening exponent. \autoref{Constitutive_Flowchart} shows the flowchart for the plastic constitutive model described above, where return mapping occurs when the flow stress exceeds the yield condition.

%------------------------------------------------------------------------------

    \subsection{Contact}
    \label{subsec2.3}

            This section introduces the contact domain method as a search algorithm to define normal and tangential (frictional) interactions between contacting bodies, specifically the structure with its plates and struts (self-contact). 
            
            When an auxetic structure undergoes a compression test, it interacts with the upper and lower plates of the testing fixture. To model the deformation of structure accurately it is essential to account for the contact phenomenon in the numerical solution. Furthermore, after reaching a certain level of strain during compression, the auxetic structure begins to densify and struts come into contact with each other which initiates self-contact. To address this type of contact in the solution process, special boundary constraints need to be imposed. 
           
           The constraints are generally defined as geometrical impenetrability and friction laws. The former is used to prevent body penetration, while the latter states the frictional behavior of the contact surfaces. Among the various approaches for the treatment of the contact problem, the contact domain theory \cite{oliver2009contact, hartmann2009contact} is used as an appropriate discretization technique for FEM encompassing the normal, tangential with friction constraints.

        \subsubsection{Contact domain method}
        \label{subsubsec2.3.1}

\begin{figure}[h]%
            \centering
            \includegraphics[width=0.4\textwidth]{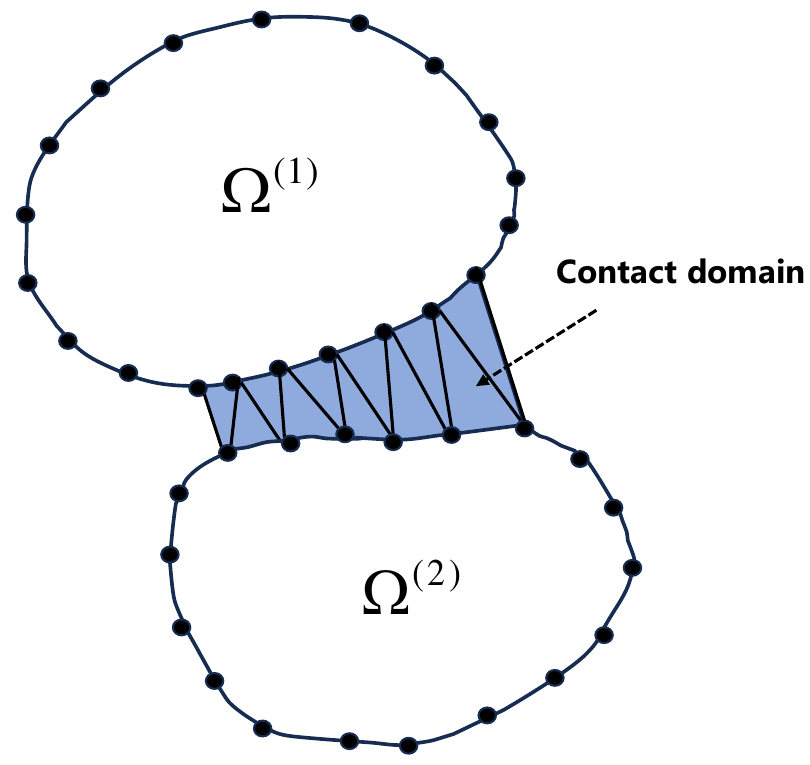}
            \caption{The patches constructed between bodies in the contact domain method.}\label{contact_domain}
            \end{figure}

            Two types of discretization techniques, node-to-segment and segment-to-segment, exist in contact description models. In these approaches, a contact point or surface is projected onto other contact surfaces, indicating a lower dimensionality of the contact problem than the bodies involved. However, the so-called contact domain provides a separate intermediate with the same dimensions to contact bodies \cite{oliver2009contact}. The contact domains are created using non-overlapping linear triangular-shaped patches (refer to \autoref{contact_domain}) implicitly identifying all potential pairings between nodes and segments and are positioned between the bodies. In contact domains, a gap, which serves as a stretch-like measure, is introduced to interpolate the displacement field from the contacting boundaries.

            The equation of the weak form (virtual work contribution) of contact employing a penalty method can be written as:
    
    	\begin{equation}\label{eq10}
                \delta {{\Pi }_{cont}}({{\mathbf{u}}^{(\alpha )}};\,\delta {{\mathbf{u}}^{(\alpha )}})=\underbrace{\int_{D_{n}^{N}}{{{{\tilde{t}}}_{N}}({{{\bar{g}}}_{N}})\delta {{{\bar{g}}}_{N}}dD}}_{normal\,\,contact}+\underbrace{\int_{D_{n}^{N}}{{{{\tilde{t}}}_{T}}({{{\bar{g}}}_{T}})\delta {{{\bar{g}}}_{T}}dD}}_{stick}+\underbrace{\int_{D_{n}^{N}/D_{n}^{T}}{\tilde{\Upsilon }\delta {{{\bar{g}}}_{T}}dD}}_{slip}\,\,\,\,\,\,\,\,\,\forall \delta {{\mathbf{u}}^{(\alpha )}}\in {{\mathbf{\nu }}_{0}}
    	\end{equation}
            where ${{\mathbf{u}}^{(\alpha )}}$ indicates the virtual displacement and $\delta {{\mathbf{u}}^{(\alpha )}}$ is its variation. ${D_{n}^{N}}$ and ${D_{n}^{T}}$ are the active normal and tangential contact domains at time $n$, respectively. $\tilde{\Upsilon }$ is the Coulomb friction law. Eq. \ref{eq10} is a function of the penalized normal ${{\tilde{t}}_{N}}$ and tangential ${{\tilde{t}}_{T}}$ directions.

\begin{figure}[h]%
            \centering
            \includegraphics[width=0.6\textwidth]{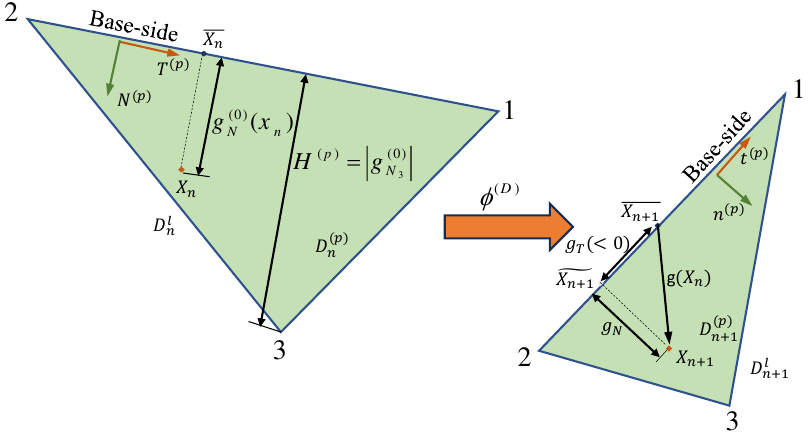}
            \caption{Linear triangle contact patch in previous (left) and current (right) configuration.}\label{Triangle_Contact}
            \end{figure}

            A triangular contact patch is depicted in \autoref{Triangle_Contact} in both the previous (left) and current (right) configurations. For every point of the contact patch ${{\mathbf{x}}_{n}}\in D_{n}^{(p)}$, a geometrical normal gap is the signed distance from its N-projection on the base side ${{\bar{x}}_{n}}\in {{\Gamma }_{D}}$ written as:
    
    	\begin{equation}\label{eq11}
                 \mathbf{g}_{n}^{(0)}({{\mathbf{x}}_{n}})=({{\mathbf{x}}_{n}}-{{\mathbf{\bar{x}}}_{n}})\cdot \mathbf{N}({{\mathbf{x}}_{n}})
    	\end{equation}        
     where vector $\mathbf{N}$ is the unit vector normal to the side placed on contacting boundaries (base side). In the current domain, the final gap vector is written as:

            \begin{equation}\label{eq12}
                \mathbf{g}({{\mathbf{x}}_{n}})={{\mathbf{x}}_{n+1}}-{{\mathbf{\bar{x}}}_{n+1}}={{\phi }^{(D)}}({{\mathbf{x}}_{n}})-{{\phi }^{(D)}}({{\mathbf{\bar{x}}}_{n}})
    	\end{equation}
     where ${{\mathbf{x}}_{n+1}}={{\phi }^{(D)}}({{\mathbf{x}}_{n}})$ and ${{\mathbf{\bar{x}}}_{n+1}}={{\phi }^{(D)}}({{\mathbf{\bar{x}}}_{n}})$ are the convected points of ${{\mathbf{x}}_{n}}$ and ${{\mathbf{\bar{x}}}_{n}}$, respectively. The final gap vector is projected onto the current normal and tangential directions, providing the normal and tangential gaps as:

            \begin{equation}\label{eq13}
                 \mathbf{g}({{\mathbf{x}}_{n}})={{g}_{N}}({{\mathbf{x}}_{n}}){{\mathbf{n}}^{(p)}}+{{g}_{T}}({{\mathbf{x}}_{n}}){{\mathbf{t}}^{(p)}}\,\,\,\Rightarrow \,\,\,\left\{ \begin{matrix}
                 {{g}_{N}}({{\mathbf{x}}_{n}})=\mathbf{g}({{\mathbf{x}}_{n}})\cdot {{\mathbf{n}}^{(p)}}  \\
                 {{g}_{T}}({{\mathbf{x}}_{n}})=\mathbf{g}({{\mathbf{x}}_{n}})\cdot {{\mathbf{t}}^{(p)}}  \\
                 \end{matrix} \right.
    	\end{equation}

            In the contact domain method, the gap in both normal and tangential directions is dimensionless, as incorporated in Eq. \ref{eq10}. So, the gap intensities are introduced as the current gaps relative to the initial normal gap as follows:

            \begin{equation}\label{eq14}
                \begin{split}
                      & {{{\bar{g}}}_{N}}({{\mathbf{x}}_{n}})=\frac{{{g}_{N}}({{\mathbf{x}}_{n}})}{|g_{N}^{(0)}({{\mathbf{x}}_{n}})|} \\ 
                     & {{{\bar{g}}}_{T}}({{\mathbf{x}}_{n}})=\frac{{{g}_{T}}({{\mathbf{x}}_{n}})}{|g_{N}^{(0)}({{\mathbf{x}}_{n}})|} \\ 
                \end{split}
            \end{equation}
        
            The gap intensities are used to define the normal and tangential contact constraints next.
    
        \subsubsection{Normal contact constraint}
        \label{subsubsec2.3.2}

            The contact force term in Eq. \ref{eq10} for the normal contact condition is written as a function of the normal gap intensity:

            \begin{equation}\label{eq15}
                {{\tilde{t}}_{N}}=K_{N}^{*}{{\bar{g}}_{N}}=K_{N}^{*}\frac{g_{N}^{(3)}}{|g_{N}^{0(3)}|}={{K}_{N}}g_{N}^{(3)}
    	\end{equation}
     where $K_{N}^{*}$ is the regularized normal penalty parameter, ${{K}_{N}}$ and $g_{N}^{(3)}$ indicate the physical normal penalty parameter and the normal component of the gap related to node "3" (see \autoref{Triangle_Contact}) which represents the classical node-to-segment gap definition, respectively. The relationship between the regularized and physical penalties is shown as:

            \begin{equation}\label{eq16}
                K_{N}^{*}=|g_{N}^{0(3)}|{{K}_{N}}=H{{K}_{N}}
    	\end{equation}
     where $H$ indicates the height of the contact patch with respect to the baseline. Finally, the normal contact conditions are as follows:

            \begin{equation}\label{eq17}
                \begin{split}
                     & {{{\tilde{t}}}_{N}}=K_{N}^{*}{{{\bar{g}}}_{N}} \\ 
                     & K_{N}^{*}=\left\{ \begin{matrix}
                     0\,\,\,\,\,\,\,\,\,\,\,\,\,\,\,\,\,\,\,\,\,\,\,\,\,\,\,if\,\,{{{\bar{g}}}_{N}}\ge 0\,\,\,\,(no\,\,contact)  \\
                     H{{K}_{N}}\,\,\,\,\,\,\,\,\,\,\,\,\,\,\,\,\,\,\,if\,\,{{{\bar{g}}}_{N}}<0\,\,\,\,(penetration)  \\
                \end{matrix} \right. \\ 
                \end{split}
            \end{equation}

        \subsubsection{Tangential contact constraint}
        \label{subsubsec2.3.2}

            For the tangential constraint, a Coulomb friction law is considered and the stick-slip condition is separated. The slip function is introduced as:
    
            \begin{equation}\label{eq19}
                \phi =\left| {{{\tilde{t}}}_{T}} \right|-\tilde{\mu} \left| {{{\tilde{t}}}_{N}} \right|\left\{ \begin{matrix}
                <0\,\,\,\,\to \,\,\,\,stick  \\
                =0\,\,\,\,\to \,\,\,\,slip\,\,\,  \\
                \end{matrix} \right.
    	\end{equation}
     where $\tilde{\mu}$ is the friction coefficient. Detailed information regarding the contact domain method can be found at \cite{oliver2009contact, hartmann2009contact}.

    \subsection{Model setup}
    \label{subsec2.4}

\begin{figure}[h!]
            \centering
            \begin{subfigure}[h]{0.3\textwidth}
                \centering
                \includegraphics[width=0.98\textwidth]{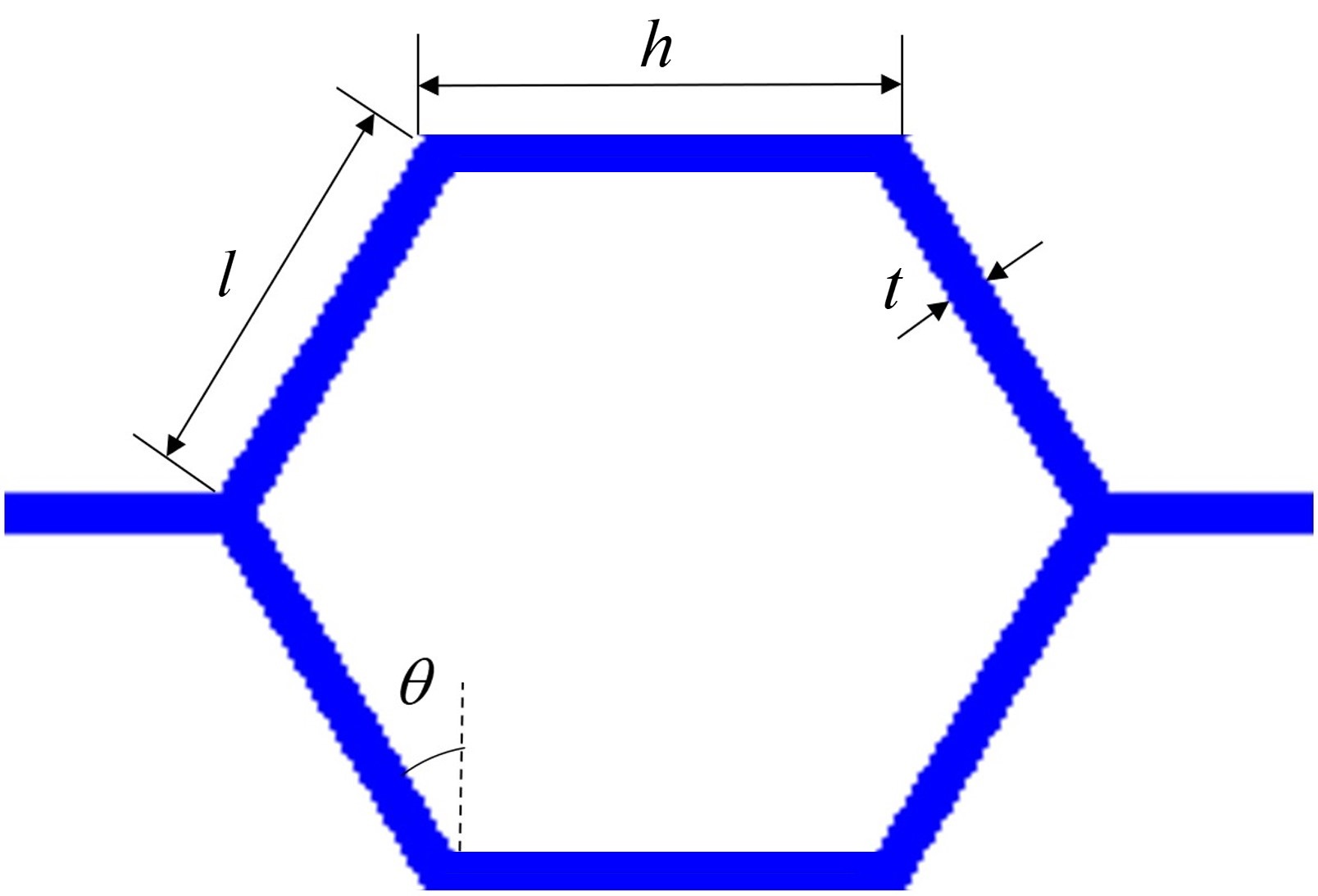}
                \vspace{0.0cm}
                \caption{}
                \label{UnitCell_Hex}
            \end{subfigure}
            \hspace{0.5cm}
            \begin{subfigure}[h]{0.3\textwidth}
                \centering
                \includegraphics[width=0.98\textwidth]{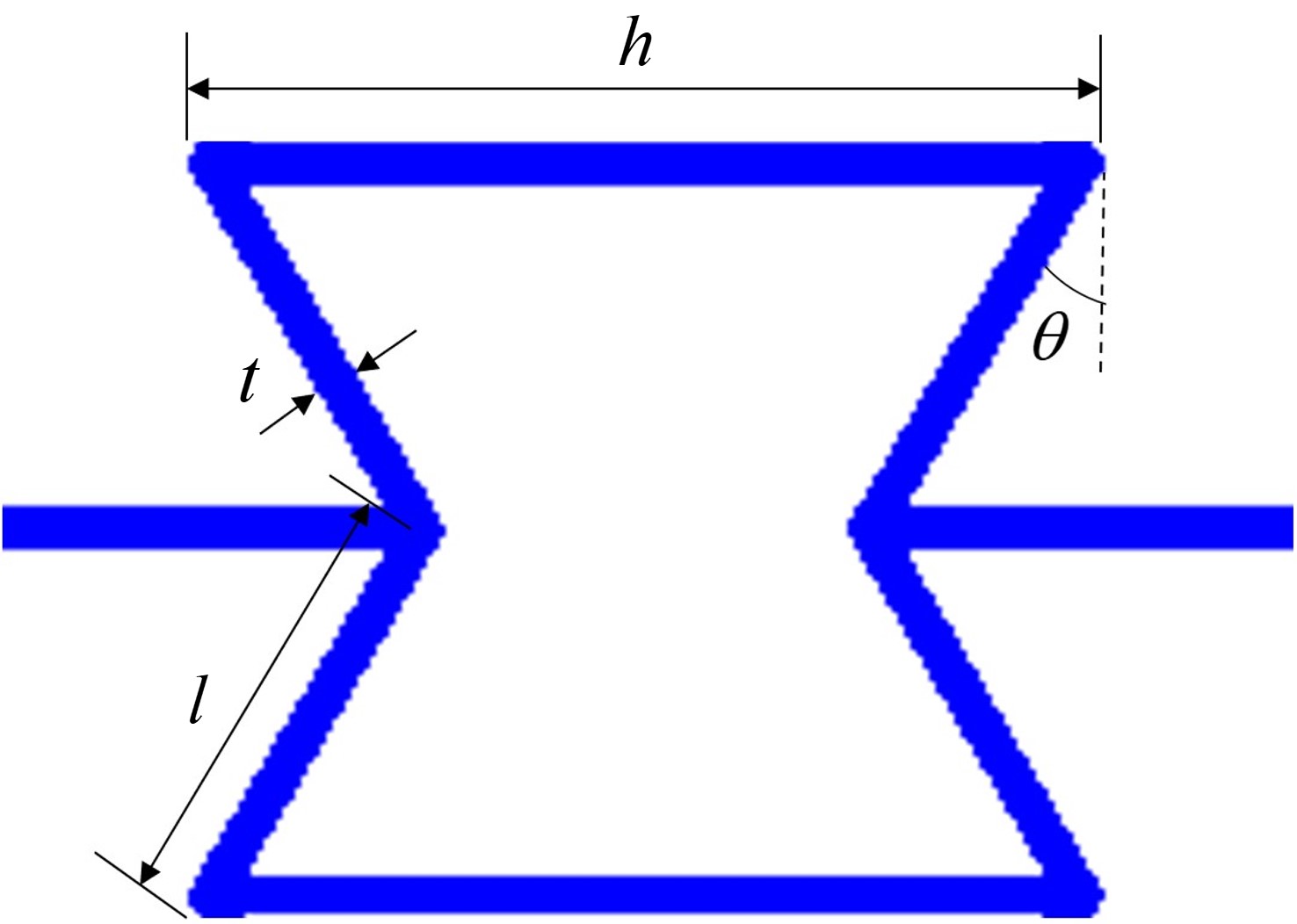}
                \caption{}
                \label{UnitCell_Re}
            \end{subfigure}
            \vspace{0.0cm}
            \caption{Geometry of the (a) HH and (b) RA unit cells.}
            \label{UnitCell}
        \end{figure}

        In the present study, the analysis focuses on HH and RA structures. The geometric characteristics of a unit cell in these structures are depicted in \autoref{UnitCell}.

        The main parameters are $h$ (the length of the horizontal struts), $l$ (the length of the inclined struts), $\theta$ (the angle of the inclined struts), and $t$ (the thickness of the struts). It should be noted that $\theta$ is positive for honeycomb and negative for re-entrant cells. \autoref{tab1} presents the design parameters of the structures.

        \begin{table}[h]
		\centering
		\fontsize{8}{13}\selectfont
		\caption{Parameters of the unit cells \cite{luo2022mechanical}.}
		\label{tab1}
		\begin{tabular}{l l l l l}
			\hline
	          & $h$ (mm)  & $t$ (mm) & $l$ (mm) & $\theta \,{{(}^{\circ }})$   \\
                \hline
        	Hexagonal cell & 14  & 1.5  & 14  & 30 \\
			Re-entrant cell  & 28  & 1.5  & 14  & 30      \\
                \hline
		\end{tabular}
        \end{table}

\begin{figure}[!h]
            \centering
            \begin{subfigure}[h]{\textwidth}
            \centering
                \begin{subfigure}[h]{0.45\textwidth}
                    \centering
                    \includegraphics[width=\textwidth]{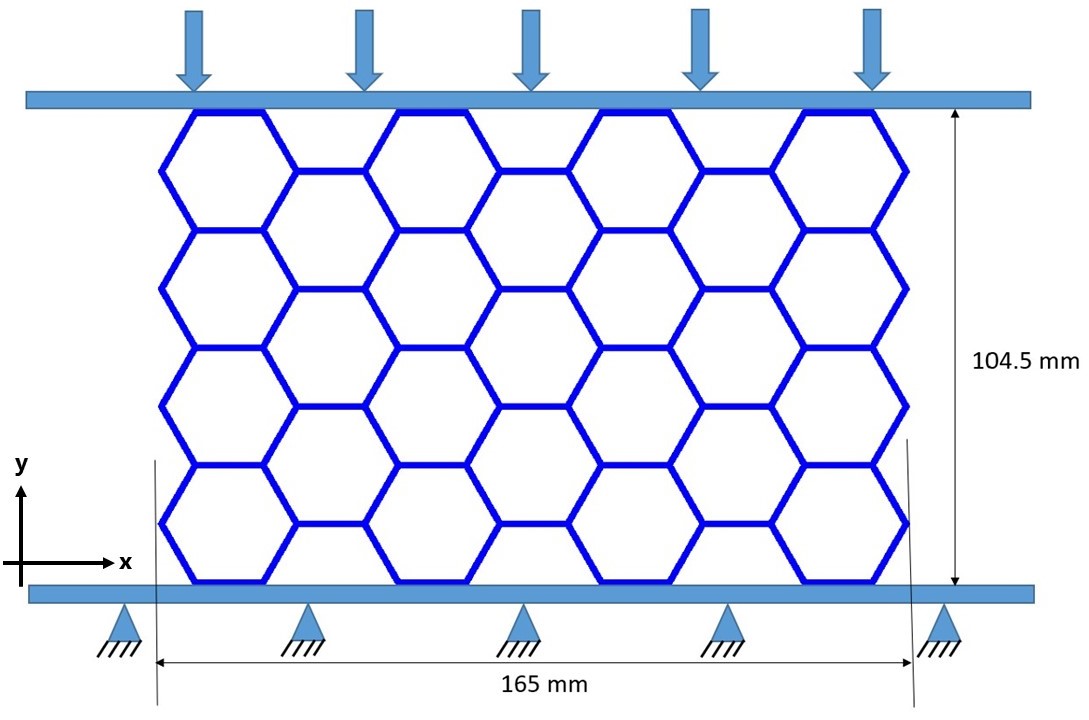}
                    \vspace{-0.5cm}
                    \caption{Hexagonal under y-directional compression}
                    \label{fig5a}
                \end{subfigure}
                \begin{subfigure}[h]{0.4\textwidth}
                    \centering
                    \includegraphics[width=\textwidth]{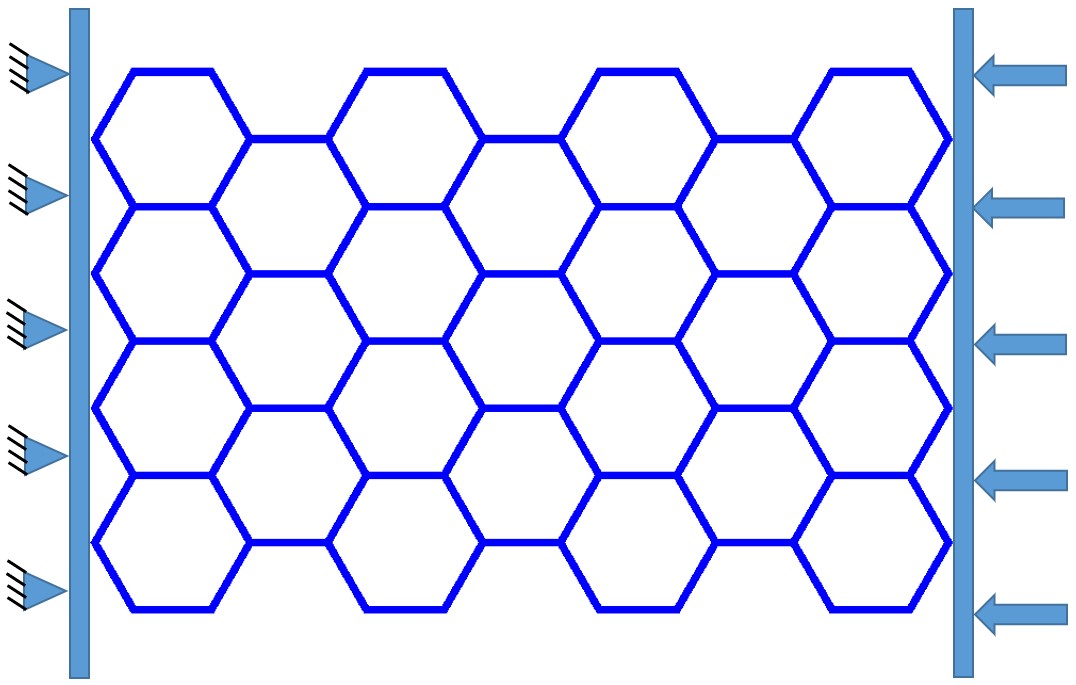}
                    \vspace{-0.5cm}
                    \caption{Hexagonal under x-directional compression}
                    \label{fig5b}
                \end{subfigure}
                \vspace{-0.1cm}
            \end{subfigure}\\
            \begin{subfigure}[h]{\textwidth}
            \centering
                \begin{subfigure}[h]{0.45\textwidth}
                    \centering
                    \includegraphics[width=\textwidth]{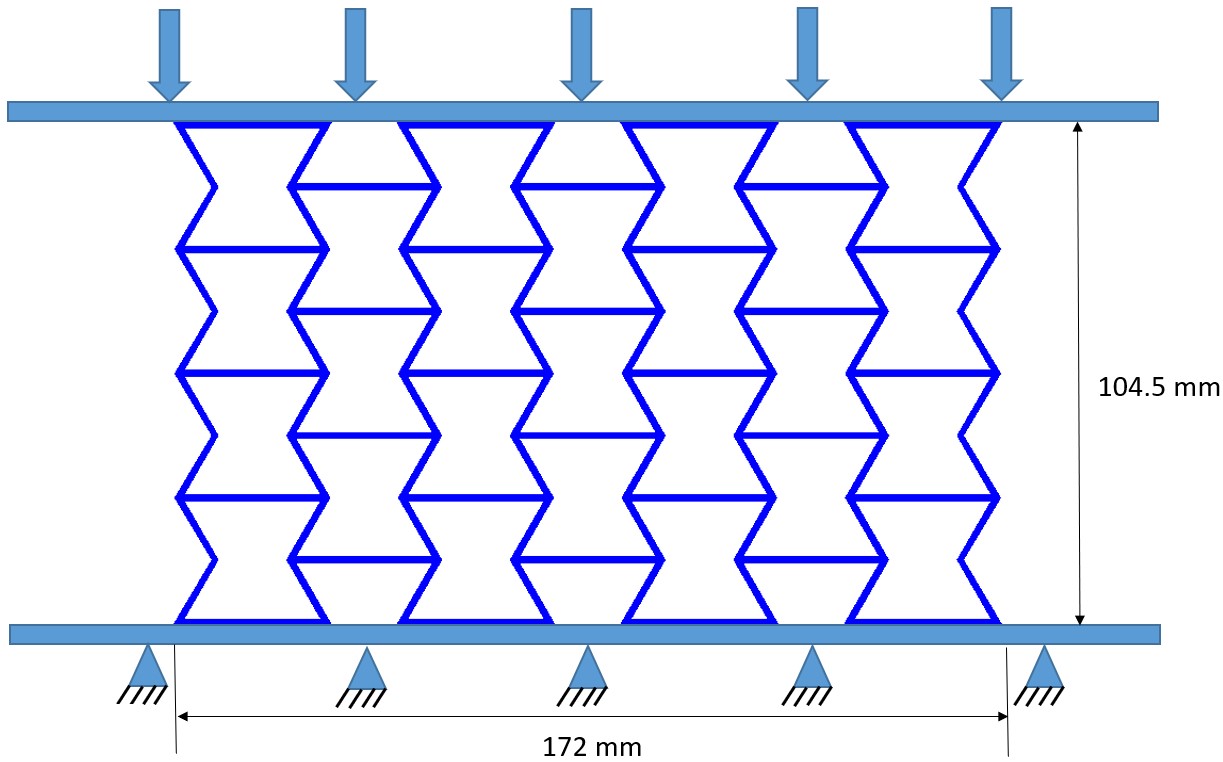}
                    \vspace{-0.5cm}
                    \caption{Re-entrant under y-directional compression}
                    \label{fig5c}
                \end{subfigure}
                \begin{subfigure}[h]{0.4\textwidth}
                    \centering
                    \includegraphics[width=\textwidth]{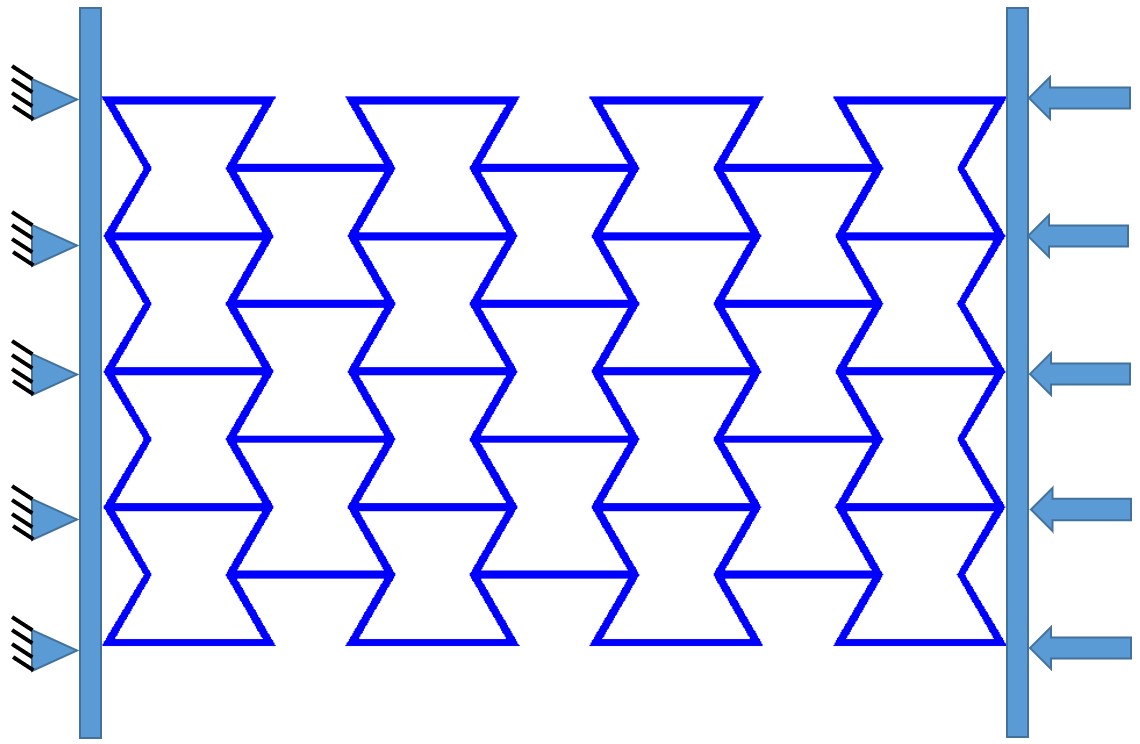}
                    \vspace{-0.5cm}
                    \caption{Re-entrant under x-directional compression}
                    \label{fig5d}
                \end{subfigure}
                \vspace{-0.1cm}
            \end{subfigure}
            \vspace{0.0cm}
            \caption{Geometry and boundary conditions of structures under compression.}
            \label{fig5}
        \end{figure}

\begin{figure}[h]%
            \centering
            \includegraphics[width=0.6\textwidth]{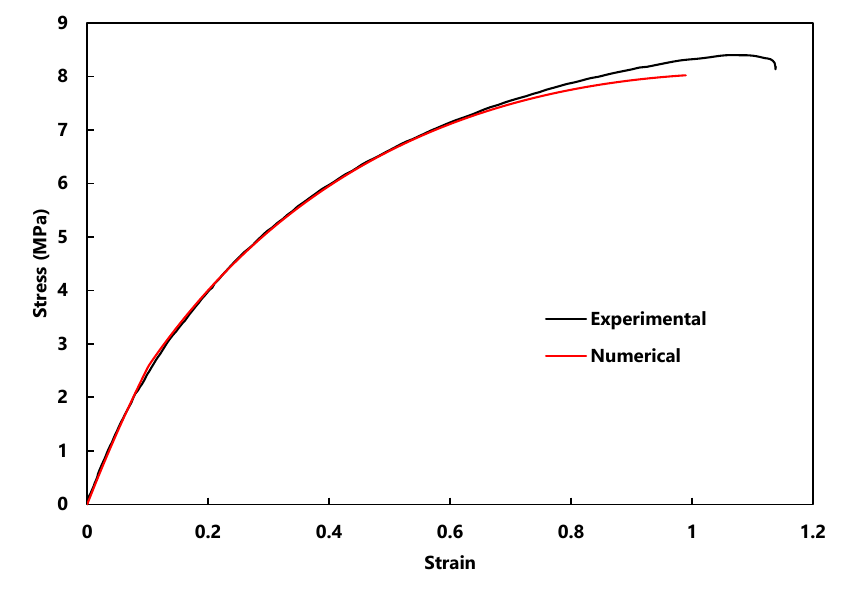}
            \caption{Stress-strain curves of uni-axial tensile test (experiments from \cite{luo2022mechanical}).}
            \label{Tensile_Test}  
        \end{figure}

        Lattice structures consist of unit cells with the same dimensions of $4\times 4$ as depicted in \autoref{fig5}. The structures are fabricated with dog bone samples of 90A TPU material. The calibration of the hyperelastic-plastic constitutive model parameters is based on the experimental data obtained in \cite{luo2022mechanical}. For the elastic part, the values of the Young's modulus, density and Poisson's ratio are, 24 MPa, 1040 kg/m\textsuperscript{3} and 0.4, respectively. The fitting result between the experiment and the numerical model is shown in \autoref{Tensile_Test} and the plastic hardening constants are listed in \autoref{tab2}.

        \begin{table}[h]
		\centering
		\fontsize{8}{13}\selectfont
		\caption{Parameters of the Simo exponential hardening law.}
		\label{tab2}
		\begin{tabular}{l l l l l}
			\hline
	           ${{\sigma }_{0}}$ (MPa)  & ${{K}_{0}}$ (MPa) & $K$ (MPa) & ${{K}_{\infty }}$ (MPa) & $n$   \\
                \hline
        	2.2 & 2  & 24  & 22  & 1.5 \\
                \hline
		\end{tabular}
        \end{table}

        Geometrical modeling and simulation are performed using GiD (pre-post processor) \cite{gidhome} and Pfem-Kratos (FE solver) \cite{PFEM}, respectively. The HH and RA structures are analyzed using both 2D plane strain and 3D with a depth of $40\,mm$. The structures are positioned between two rigid plates. The bottom plate is fully constrained while the top plate is free to translate in the y direction and displacement is applied to the top plate (\autoref{fig5a} and \autoref{fig5c}). Similarly, along the x-direction, the left plate is fully constrained, and displacement is applied to the right plate (\autoref{fig5b} and \autoref{fig5d}). Also, the boundary conditions for the tensile-loaded cases are similar. Linear triangular and tetrahedral U-P hybrid elements, with a converged size of 0.2 mm, are used to discretize the models in 2D and 3D, respectively. Contact between the structures and the plates, as well as self-contact within the structures, is considered in the models. In the tangential contact, the friction coefficient is $0.4$ (\cite{luo2022mechanical}).

 %------------------------------------------------------------------------------
 %------------------------------------------------------------------------------

    \section{Results and discussion}
    \label{sec3}
    
        In this section, the mechanical response of honeycomb and auxetic re-entrant structures under compression and tension are studied. Their deformation response, equivalent Poisson's ratio and energy absorption are investigated. The aim is to understand how different loading directions affect the performance of RA and HH structures.

	\subsection{Deformation analysis}
	\label{subsec3.1}

            \subsubsection{Compression y-direction}
            \label{subsubsec3.1.1}

            \noindent\textit{3.1.1.1} 2D analysis

\begin{figure}[!]
            \centering
            \begin{subfigure}[h]{0.5\textwidth}
                    \centering
                    \includegraphics[width=\textwidth]{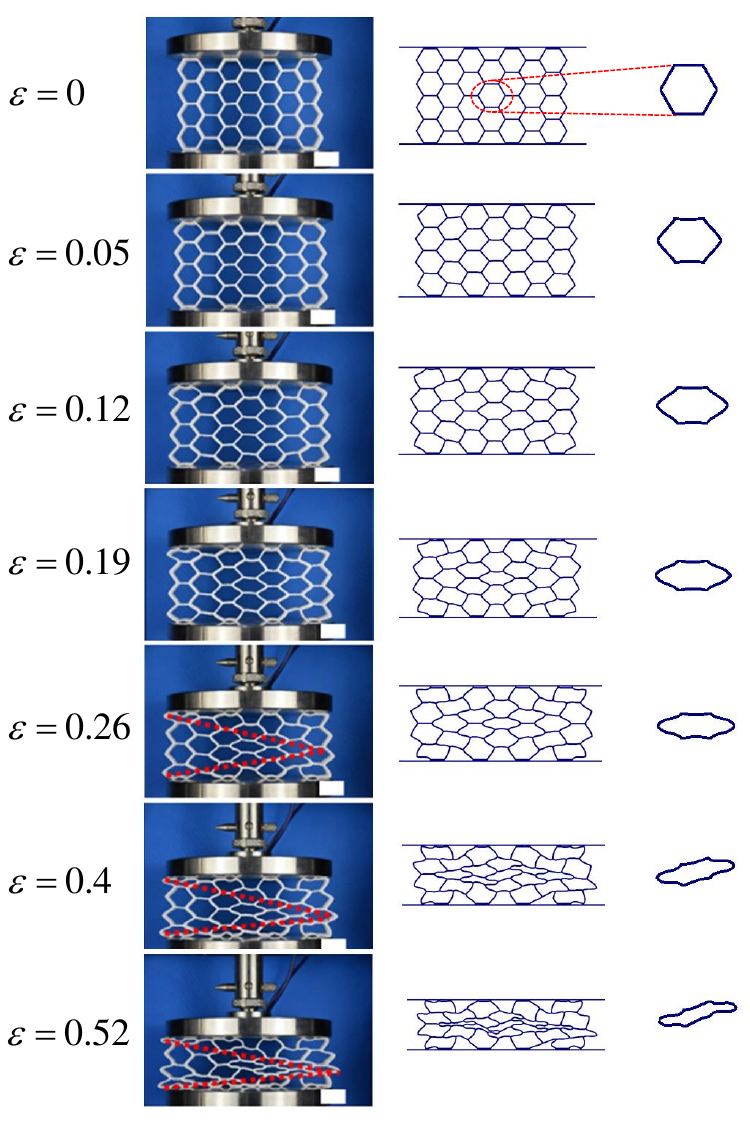}
                    \vspace{-0.5cm}
                    \caption{}
                    \label{fig7a}
                \vspace{-0.1cm}
            \end{subfigure}\\
            \begin{subfigure}[h]{0.6\textwidth}
                    \centering
                    \includegraphics[width=\textwidth]{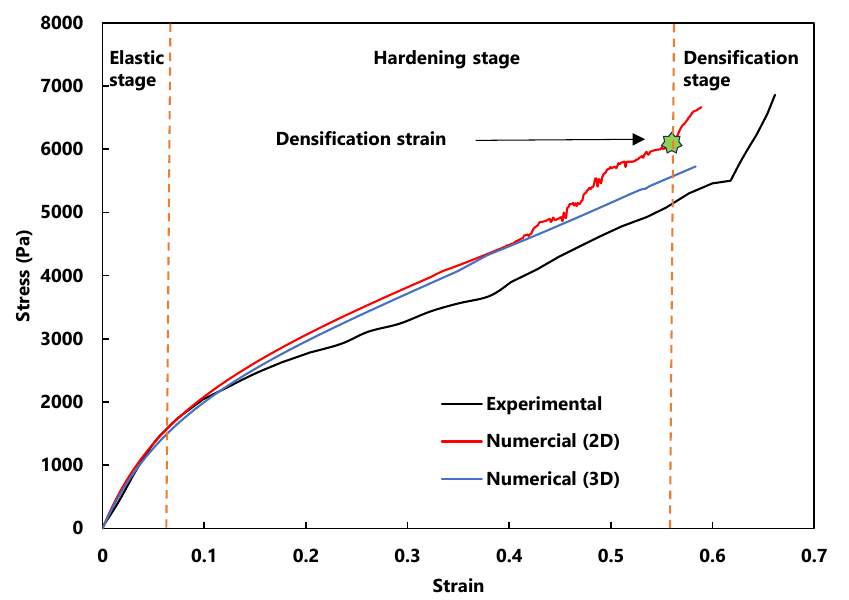}
                    \vspace{-0.5cm}
                    \caption{}
                    \label{fig7b}
                \vspace{-0.1cm}
            \end{subfigure}
            \vspace{0.0cm}
            \caption{(a) Deformation pattern and (b) stress-strain curves of the HH structure under compression in the y-direction (experiments from \cite{luo2022mechanical}).}
            \label{fig7}
            \end{figure}

\begin{figure}[!]
            \centering
            \begin{subfigure}[h]{0.5\textwidth}
                    \centering
                    \includegraphics[width=\textwidth]{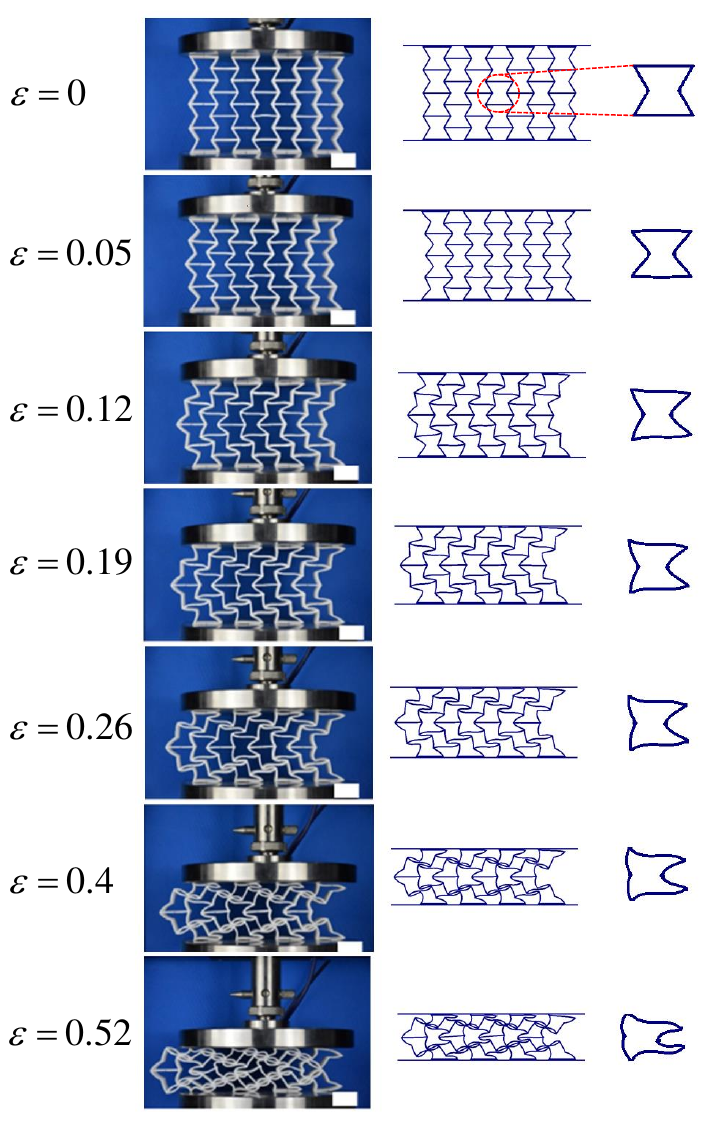}
                    \vspace{-0.5cm}
                    \caption{}
                    \label{fig8a}
                \vspace{-0.1cm}
            \end{subfigure}\\
            \begin{subfigure}[h]{0.6\textwidth}
                    \centering
                    \includegraphics[width=\textwidth]{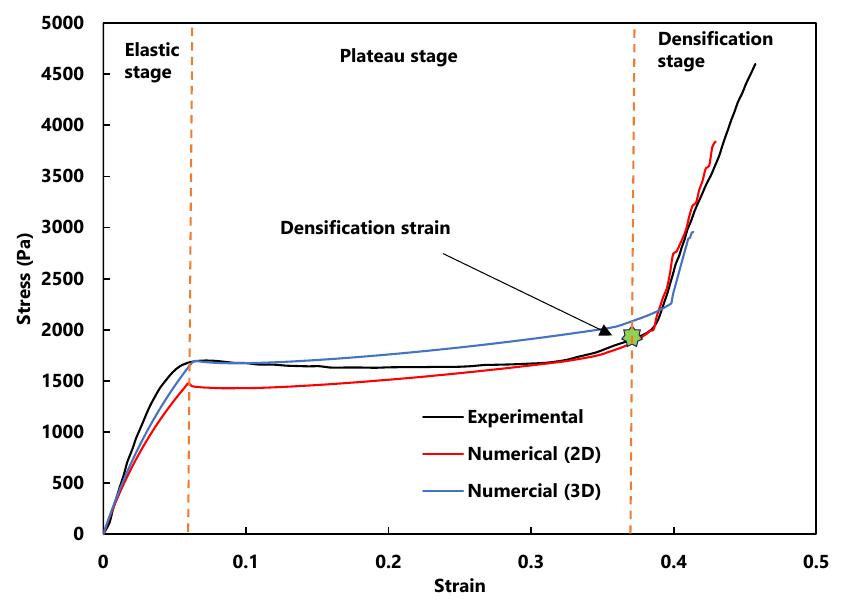}
                    \vspace{-0.5cm}
                    \caption{}
                    \label{fig8b}
                \vspace{-0.1cm}
            \end{subfigure}
            \vspace{0.0cm}
            \caption{(a) Deformation pattern and (b) stress-strain curves of the RA structure under compression in the y-direction (experiments from \cite{luo2022mechanical}).}
            \label{fig8}
            \end{figure}

            The deformation patterns and stress-strain curves of honeycomb and re-entrant structures under compression in the y-direction at increasing strain levels are presented in \autoref{fig7} and \autoref{fig8}, respectively.

         In a honeycomb structure, initially, the cell walls undergo elastic buckling (\autoref{fig7a} at strain $\varepsilon =0.05$), followed by plastic deformation where strain hardening occurs. The plastic strain hardening stage begins at strain $\varepsilon =0.065$ and continues up to strain $\varepsilon =0.56$ (\autoref{fig7b}). Once the structure reaches the densification strain i.e., $\varepsilon =0.56$, the cell walls fold and stack on top of each other, resulting in the densification of the structure. The trend of cell folding in the deformation is unpredictable and is associated with the type of structure. From \autoref{fig7a} it can be seen that the HH exhibits a reclining v-shaped deformation pattern \cite{luo2022mechanical}. Following the numerical deformation patterns, it can be seen that the distance between the top and bottom edge of the unit cell is decreased and the deformed shape is altered significantly. Densification occurs when the edges of the unit cells are in contact at higher strains. Both the deformation pattern and stress-strain curve of the HH structure were compared to available experimental results, showing good agreement (\autoref{fig7}). The deviation between experimental and numerical results in predicting stress reaches a maximum level of approximately 18\% during the hardening stage.   
         
         In the re-entrant structure, likewise, the cell walls initially undergo elastic buckling (elastic stage at \autoref{fig8a}) up to strain $\varepsilon =0.06$, then the cell walls rotate and fold over each other. The folding and rotating conditions depend on geometrical properties such as the thickness or lengths of the cell walls. The stress-strain curve can be divided into three stages: the linear elastic, plateau, and densification \autoref{fig8b}. In the first stage, stress increases at a constant rate with increasing strain up to $\varepsilon =0.06$. In the subsequent stage, which begins with the rotation of the cell walls, stress remains nearly constant while strain continues to increase. This is known as the plateau stage and is required to determine energy absorption. In \autoref{fig8b}, it can be observed that the plateau stage starts at a strain of $\varepsilon =0.06$ and ends at a strain of approximately $\varepsilon =0.37$. During this period, the unit cells only undergo rotation (in \autoref{fig8a} from $\varepsilon =0.12$ to $\varepsilon =0.26$). This coincides with the stress-strain graph where the stress does not change by the strain variation. The final stage is densification, in which the cell walls come into contact with each other, subsequently, the stress once again increases as strain continues to rise. Once the structure reaches the densification strain i.e., $\varepsilon =0.37$, the cells compact together, making the structure stiffer (densification stage in \autoref{fig8b}). The strain-hardening phenomenon occurs in this stage. Note that in honeycomb structures, after the linear elastic stage, there is no plateau stage, and the stress continues to increase. Then, on the densification strain point, the densification stage starts. Upon the onset of densification, the structure begins to collapse inward, causing the material to densify and resulting in a rapid increase in stiffness. Comparing the experimental and numerical results for both the deformation pattern and stress-strain curve of the re-entrant structure reveals good agreement. The general shapes of the simulated structure at different stages (strains) are similar to the tested structure (\autoref{fig8a}). The predicted values of stress at plateau and densification stages closely match the experimental results. However, some minor discrepancies can be observed during the elastic stage (\autoref{fig8b}). The experimental stress values exceed the numerical predictions until a strain of $\varepsilon = 0.3$ is reached. These differences can be attributed to several simplifications in the numerical process, including the consideration of an estimated frictional coefficient between plates and structures, the use of simple linear triangular elements, and semi-accurate mesh refinement to minimize excessive computational costs.        

        \noindent\textit{3.1.1.2} 3D analysis

\begin{figure}[h]%
            \centering
            \includegraphics[width=0.7\textwidth]{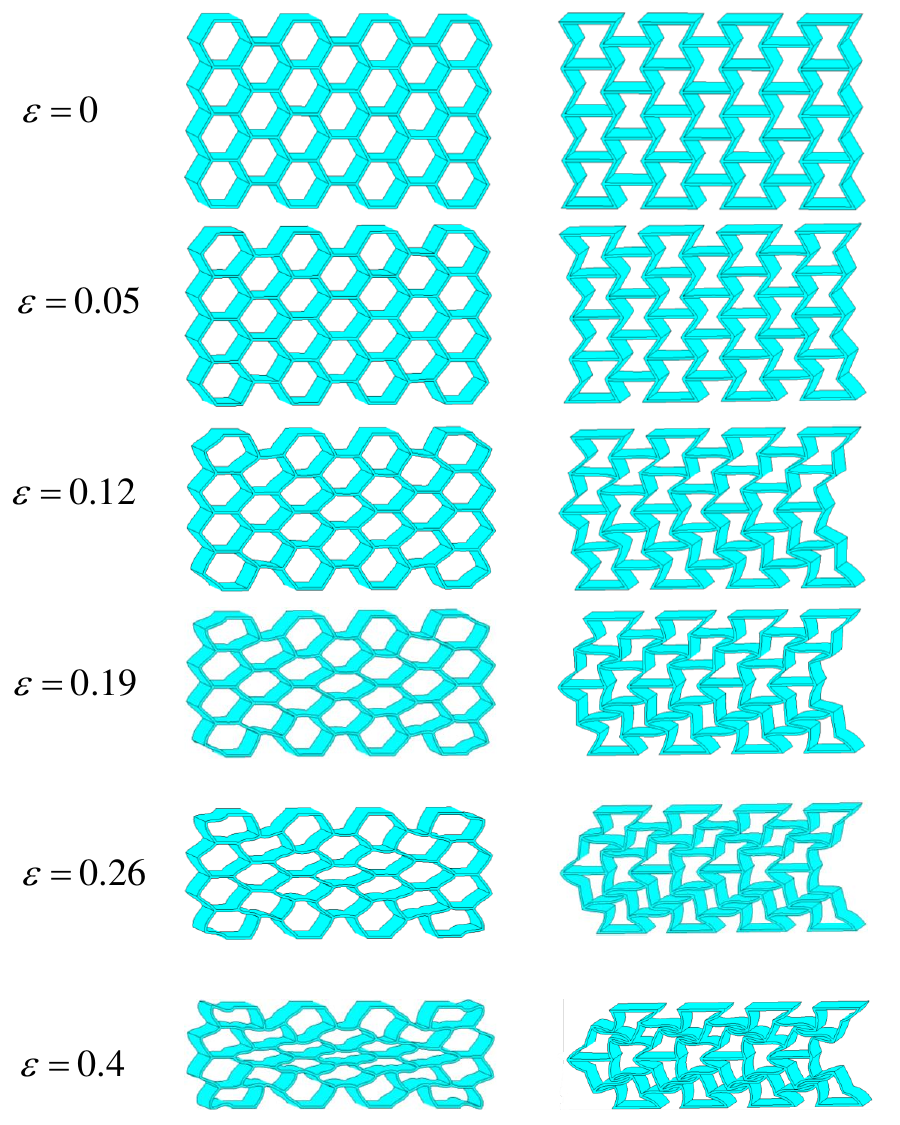}
            \caption{Deformation pattern of HH (left) and RA (right) 3D structures.}
            \label{3D_pattern}  
        \end{figure}

            The previously mentioned results for compression loading are obtained under a 2D plane strain assumption. To analyze a structure using plane strain theory, the through-thickness strain should be negligible. This condition holds for thick structures in which the thickness is significantly larger compared to its other dimensions. However, the thickness of the structures in this study is smaller than its in-plane dimensions. To ensure the validity of the plane strain assumption, 3D HH and RA structures under y-directional compression models are simulated. From \autoref{fig7b}, it is evident that the stress-strain curve of the HH in the 3D analysis closely resembles the experiment as well as the numerical results obtained for 2D cases. During the hardening stage, there are some minor differences between the stress values in the 2D and 3D responses. This can be justified by the negligible out-of-plane deformation resulting from in-plane loading in the 3D analysis. \autoref{fig8b} presents the stress-strain curve for RA structures in 3D. Likewise, 3D results show good agreement with experimental and 2D numerical results. The deformation patterns of the HH and RA 3D structures are exhibited in \autoref{3D_pattern}. For both structures, similar deformed shapes can be observed at different strains when compared with the experimental and 2D numerical results (\autoref{fig7a} and \autoref{fig8a}). Furthermore, in 3D deformed structures, the through-thickness strain reaches a maximum value of $\varepsilon = 0.03$ for the inner unit cells. This level of out-of-plane strain can be disregarded, allowing us to continue with a 2D plane strain assumption, which leads to significantly reduced computational cost. Therefore, the following analyses are conducted using a 2D plane strain assumption.

        \subsubsection{Compression x-direction}
            \label{subsubsec3.1.2}

\begin{figure}[h!]
            \centering
            \begin{subfigure}[h]{0.48\textwidth}
                    \centering
                    \includegraphics[width=\textwidth]{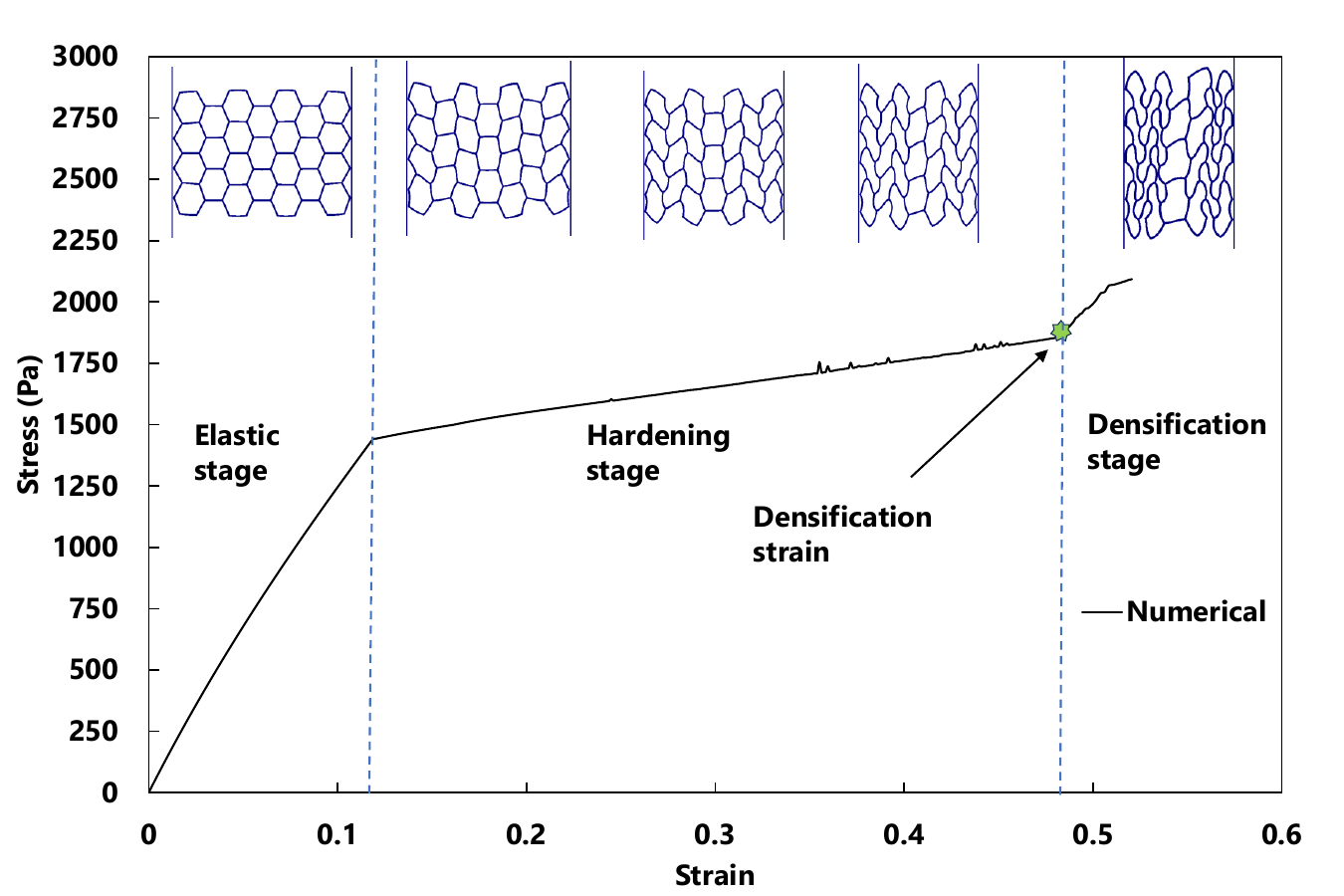}
                    \vspace{-0.5cm}
                    \caption{}
                    \label{fig9a}
                \vspace{-0.1cm}
            \end{subfigure}
            \begin{subfigure}[h]{0.49\textwidth}
                    \centering
                    \includegraphics[width=\textwidth]{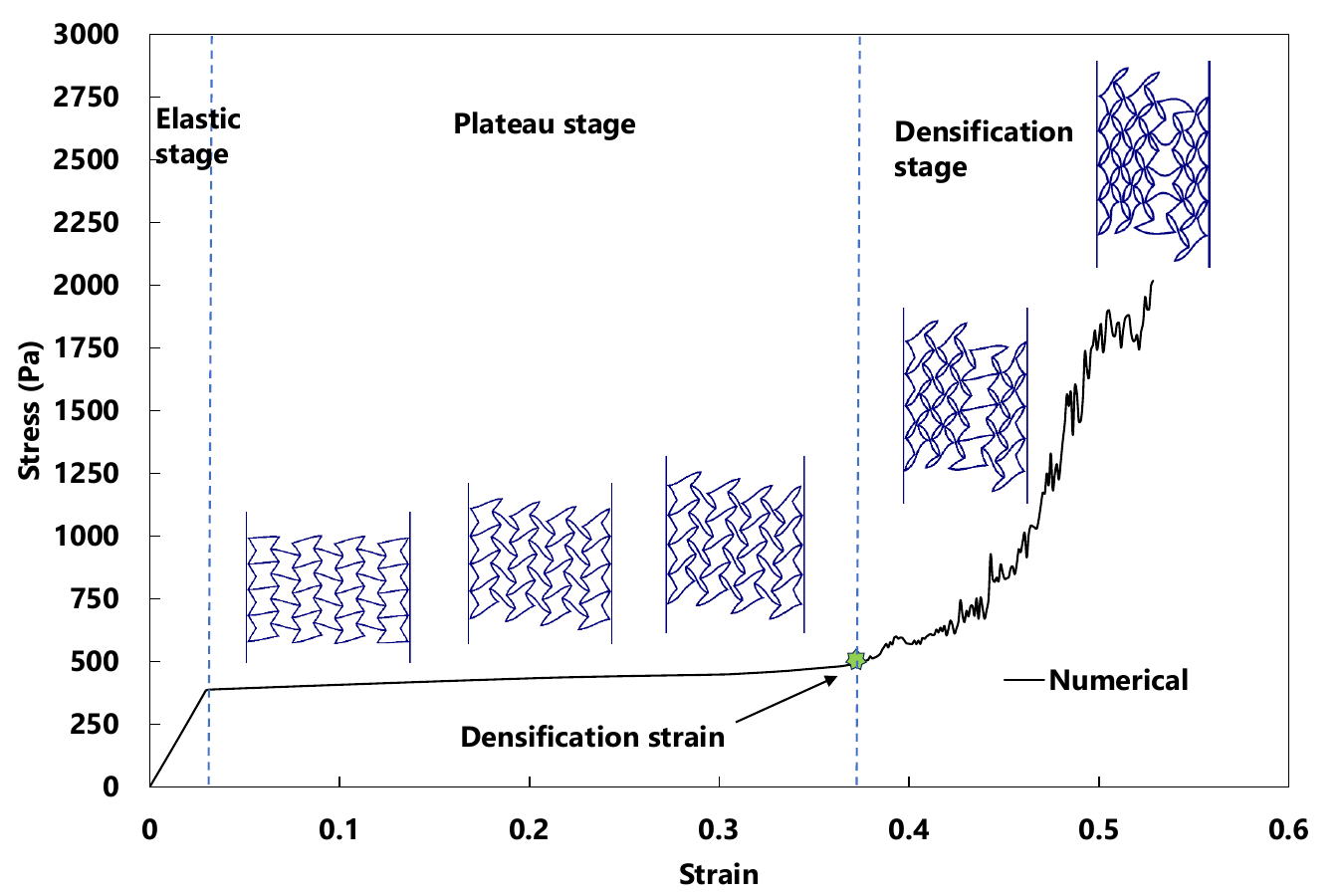}
                    \vspace{-0.5cm}
                    \caption{}
                    \label{fig9b}
                \vspace{-0.1cm}
            \end{subfigure}
            \vspace{0.0cm}
            \caption{Stress-strain curves with deformation patterns of (a) HH and (b) RA structures under compression in the x-direction.}
            \label{fig9}
        \end{figure}

         The stress-strain curves for honeycomb and re-entrant structures together with their deformation patterns at different stages of compression in the x-direction are shown in \autoref{fig9a} and \autoref{fig9b}, respectively. The curves are similar for compression in the y-direction exhibiting the same deformation stages. However, the stress levels in the y-direction loading are greater than those in the x-directional compression for both structures. This is consistent with the fact that the orientation of the unit cells influences the structural strength. In y-directional compression, the struts of the unit cells in both structures allow for a continuous transfer of the loading to the bottom support. In contrast, in x-directional compression, the orientation of the unit cells results in a discontinuous load transfer, as the struts are not aligned in the loading direction continuously. Consequently, the structural stiffness along the y-direction is higher, indicating their superior performance in that direction. In \autoref{fig9a}, the elastic stage of the hexagonal structure is larger than that of y-directional compression where the initiation strain of the hardening exceeds $\varepsilon =0.1$. Nevertheless, the hardening stage up to the densification start point becomes shorter and the slope of the curve is less steep. Therefore, the hardening rate is lower. Also, the densification strain i.e., $\varepsilon =0.48$, is lower than that of y-directional compression ($\varepsilon =0.56$). This indicates that the hexagonal structure exhibits greater strength under compression loading in the y-direction, and the duration of the hardening stage is longer when compared to compression in the x-direction. Furthermore, the deformation patterns of the hexagonal structure under compression in the x-direction (\autoref{fig9a}) show a significant difference compared to those under y-directional compression (\autoref{fig7a}). In the former, the inner cells deform less, and the outer cells predominantly deform and change shape. Conversely, in the latter, the inner unit cells undergo significant deformation, while the outer ones exhibit less deformation, consistent with experimental results. This suggests distinct deformation behaviors for the HH structure when subjected to different loading directions. It can be attributed to the contact length of the structures with plates. In x-directional compression, the outer unit cells of the structures make contact with the plates at points (edges in 3D) instead of edges (surfaces in 3D). They are compressed initially until the contact length increases, after which the inner cells begin to deform. This is contrary to the y-directional compression, where the edges of the unit cells are in contact with plates, resulting in greater strength of the outer cells in comparison to inner cells.   

         In \autoref{fig9b}, the plateau stage of the re-entrant structure under compression in the x-direction starts at $\varepsilon =0.03$ which is lower than that of y-directional compression. The plateau stage is slightly longer; however, the densification strain is almost identical. The deformation patterns in x-directional compression for the re-entrant structure differ from that in y-directional loading. During the plateau stage, all unit cells of the re-entrant structure in y-directional compression rotate in the same direction (\autoref{fig8a}); but, in x-directional compression, rows of unit cells exhibit opposite rotational behavior (\autoref{fig9b}). Several rows of unit cells near the left plate rotate upwards; however, some rows near the right plate rotate downward. Also, there is one row of unit cells that does not rotate significantly.

        \subsubsection{Tension}
            \label{subsubsec3.1.3}

\begin{figure}[h!]
                \centering
                \begin{subfigure}[h]{0.49\textwidth}
                        \centering
                        \includegraphics[width=\textwidth]{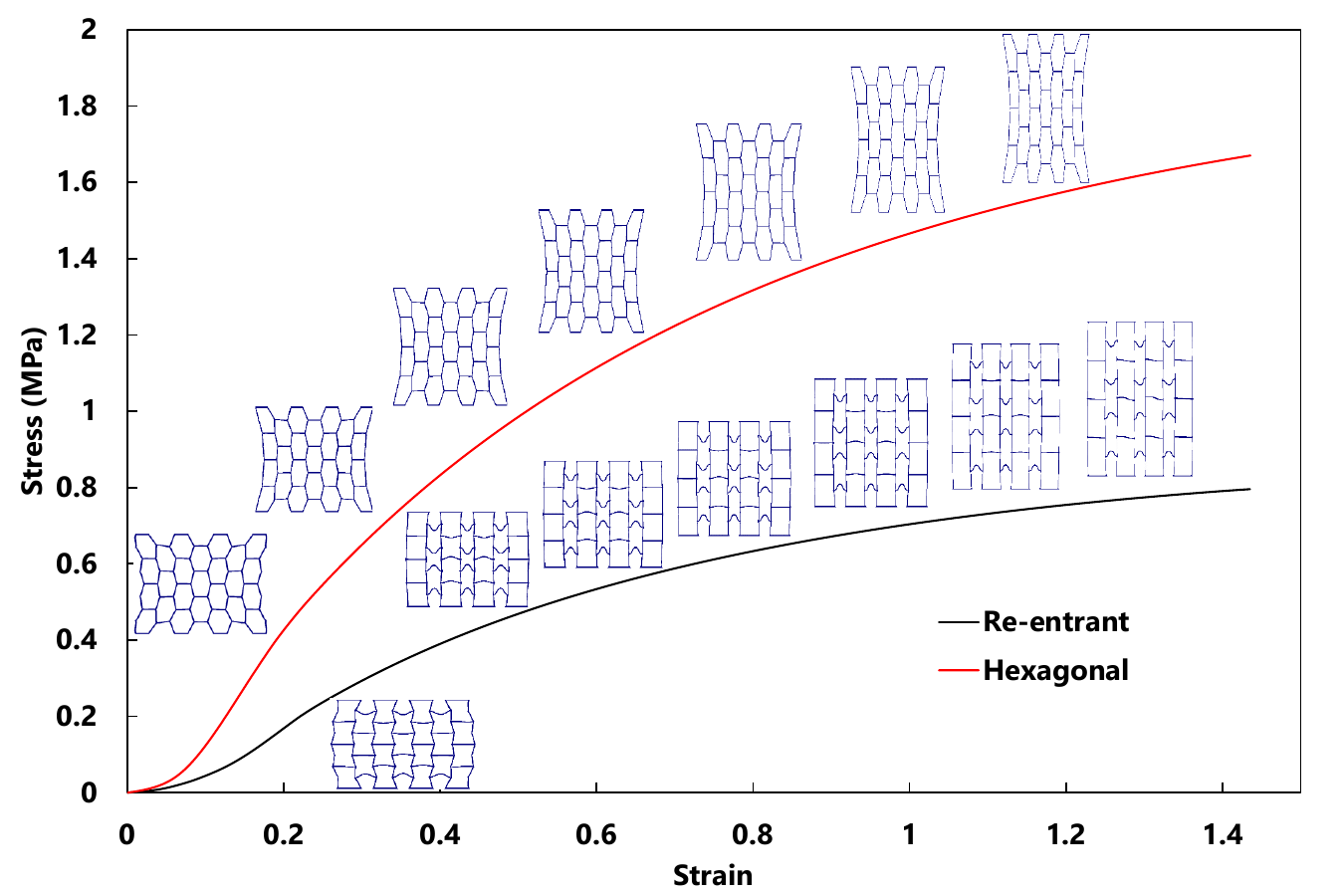}
                        \vspace{-0.5cm}
                        \caption{}
                        \label{fig10a}
                    \vspace{-0.1cm}
                \end{subfigure}
                \begin{subfigure}[h]{0.49\textwidth}
                        \centering
                        \includegraphics[width=\textwidth]{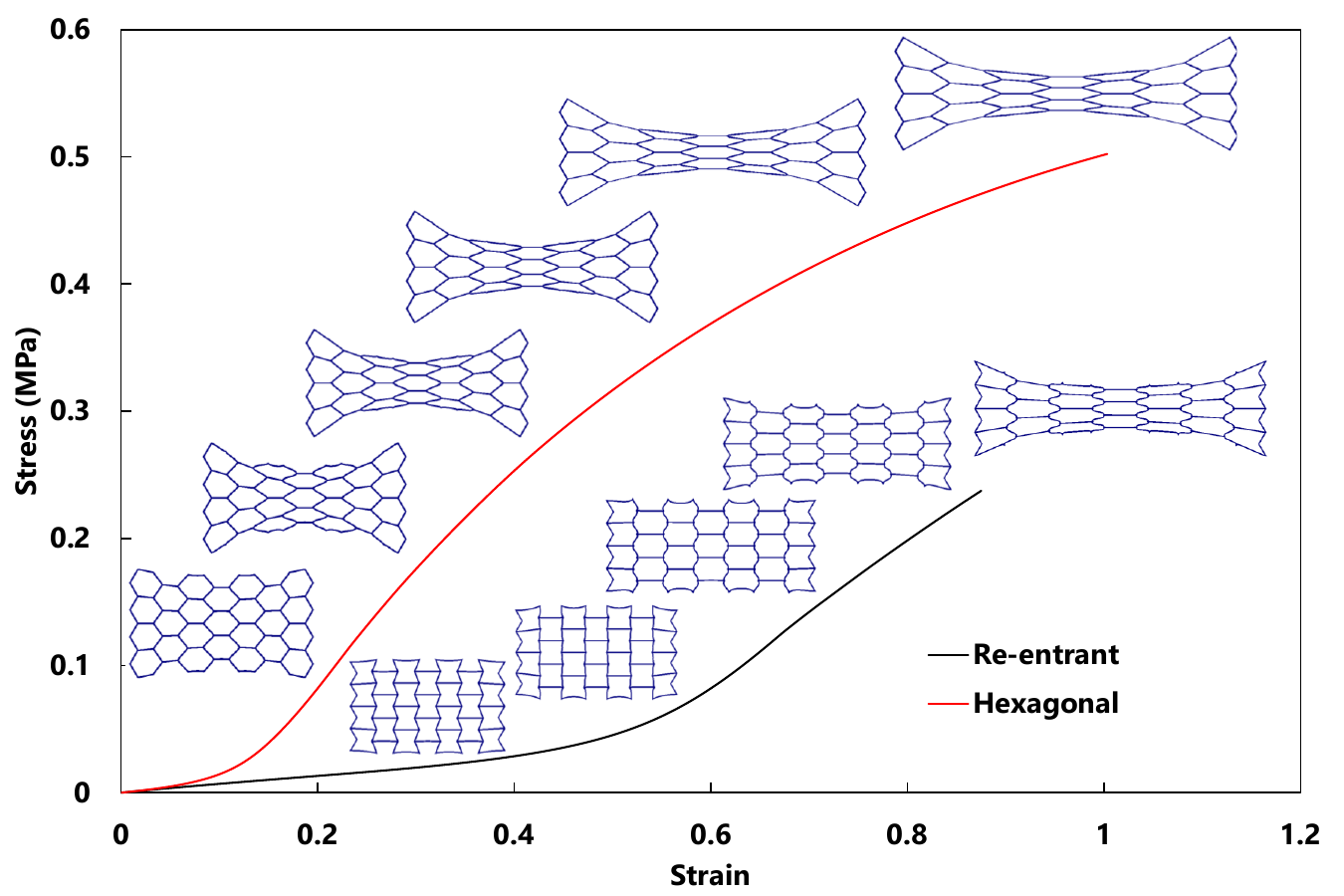}
                        \vspace{-0.5cm}
                        \caption{}
                        \label{fig10b}
                    \vspace{-0.1cm}
                \end{subfigure}
                \vspace{0.0cm}
                \caption{Stress-strain curves and deformation patterns of re-entrant and HH structures under (a) y and (b) x directional tension.}
                \label{fig10}
            \end{figure}

            In this section, the HH and RA structures are analyzed under tensile loading in both the x and y directions. The stress-strain curves and deformation patterns are presented in \autoref{fig10a} and \autoref{fig10b}. It can be seen that in both loading directions, there are no distinct stages. In both structures, the stress increases continuously as the strain grows. Therefore, the hardening occurs throughout the entire period of tensile loading. Similar to compressive loading, the structures are stiffer in the y-direction compared to the x-direction tensile loading. Hence, the stress levels in the former are greater than those in the latter. The structures in the x-directional loading experience more stretching, resulting in greater deformation changes compared to the y-direction tension. Under x-directional tension, the structure notably reduces in dimension along the y-direction (see \autoref{fig10b}). This directly influences the value of the equivalent Poisson’s ratio which is discussed next.

%------------------------------------------------------------------------------

	\subsection{Equivalent Poisson's ratio}
	\label{subsec3.2}

\begin{figure}[h!]
                \centering
                \begin{subfigure}[h]{0.49\textwidth}
                        \centering
                        \includegraphics[width=\textwidth]{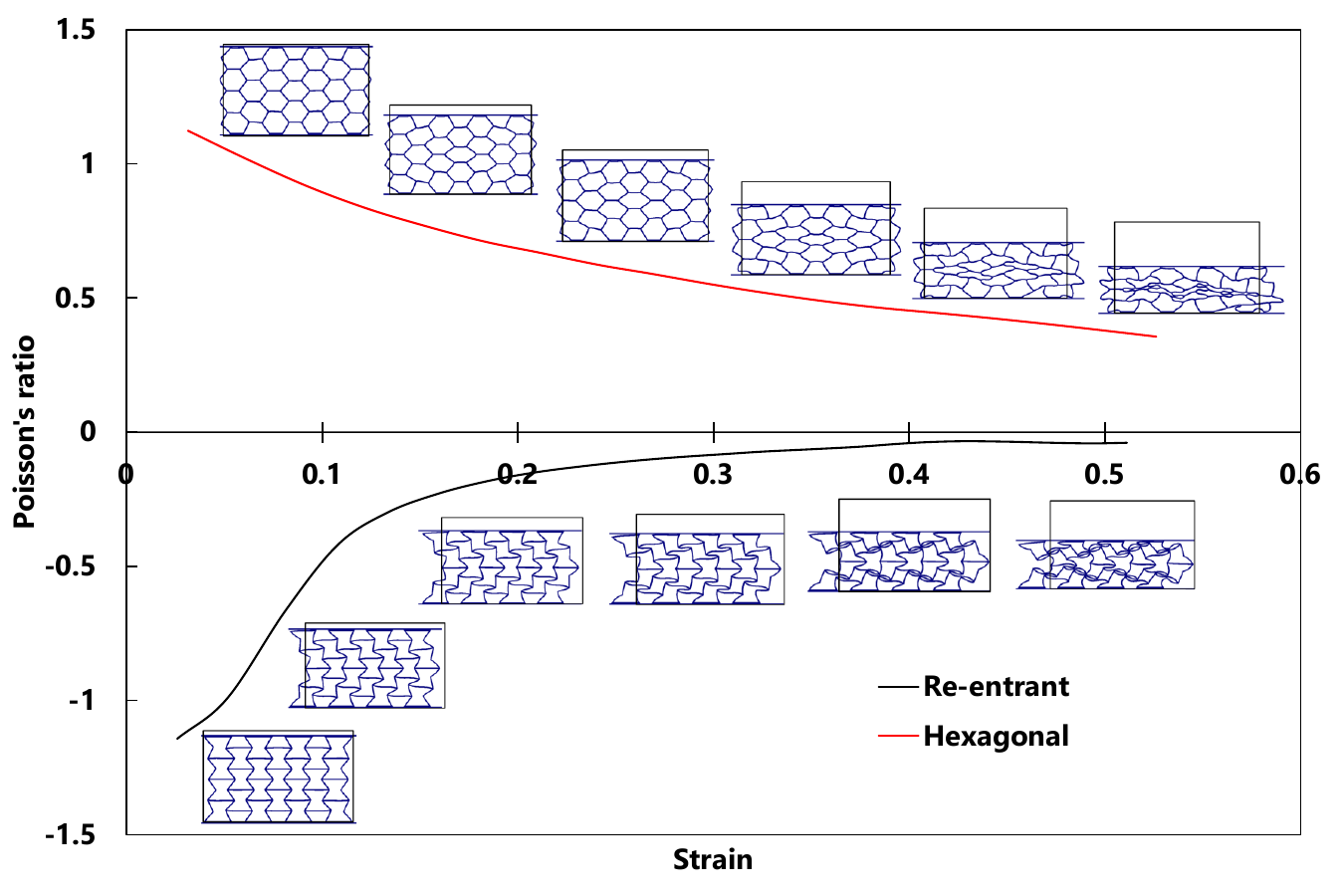}
                        \vspace{-0.5cm}
                        \caption{}
                        \label{fig11a}
                    \vspace{-0.1cm}
                \end{subfigure}
                \begin{subfigure}[h]{0.49\textwidth}
                        \centering
                        \includegraphics[width=\textwidth]{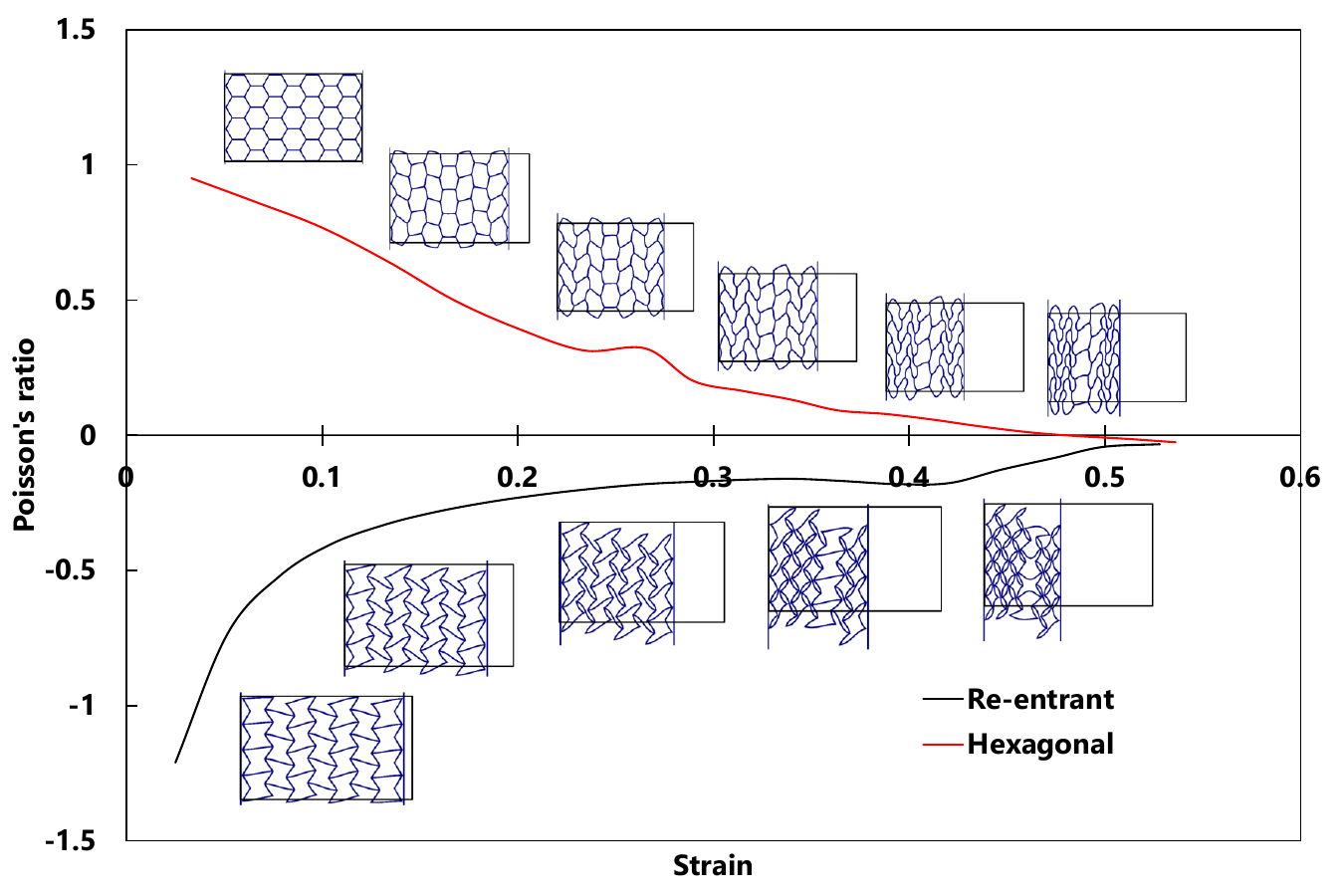}
                        \vspace{-0.5cm}
                        \caption{}
                        \label{fig11b}
                    \vspace{-0.1cm}
                \end{subfigure}
                \vspace{0.0cm}
                \caption{Poisson's ratios of RA and HH structures under (a) y and (b) x directional compression.}
                \label{fig11}
            \end{figure}

           \autoref{fig11a} and \autoref{fig11b} show the equivalent Poisson's ratio of both lattice structures under compression loading in the y and x directions, respectively. The deformation patterns of the structures subjected to the compression are also displayed along the curves, demonstrating expansion and contraction laterally for the honeycomb and re-entrant structures. For both loading directions, the Poisson's ratio remains negative for the RA structure while it is positive for the honeycomb structure as the loading strain increases. This highlights that the re-entrant structure exhibits auxetic behavior for both x and y compression loading directions. In both structures, Poisson's ratio tends to zero as the strain increases, indicating a gradual weakening of the lateral effect. As seen in \autoref{fig11}, the Poisson's ratio of the hexagonal structure decreases continuously and when the strain reaches $\varepsilon =0.52$ the Poisson's ratio becomes 0.35. However, in the x-directional compression, it reaches zero at the same strain (\autoref{fig11b}). This implies that under large strains during compression along the x-direction, the hexagonal structure does not expand. Rather, at specific strain thresholds, it contracts, leading to the marginal negative values in the Poisson's ratios.

\begin{figure}[h!]
                \centering
                \begin{subfigure}[h]{0.49\textwidth}
                        \centering
                        \includegraphics[width=\textwidth]{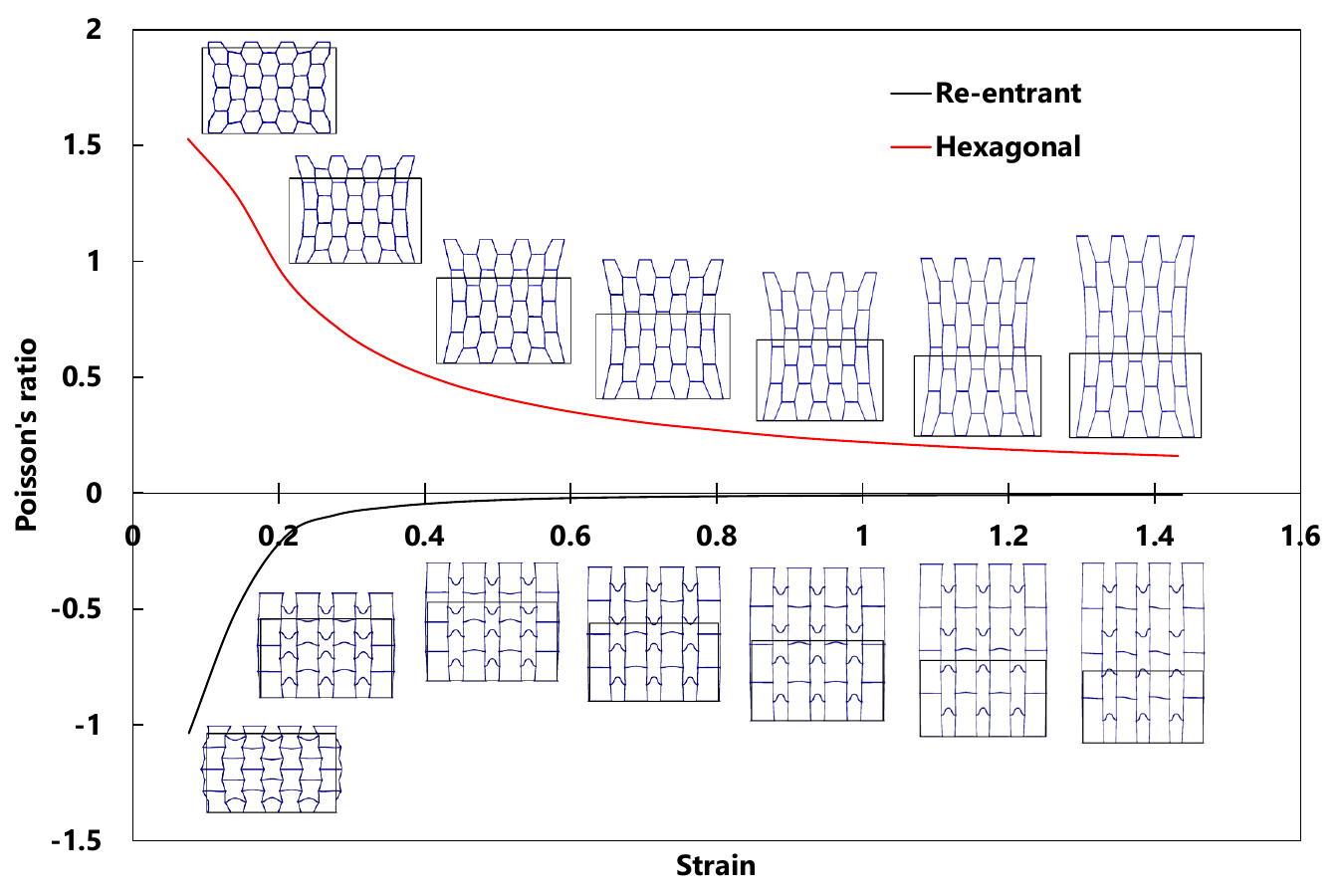}
                        \vspace{-0.5cm}
                        \caption{}
                        \label{fig12a}
                    \vspace{-0.1cm}
                \end{subfigure}
                \begin{subfigure}[h]{0.49\textwidth}
                        \centering
                        \includegraphics[width=\textwidth]{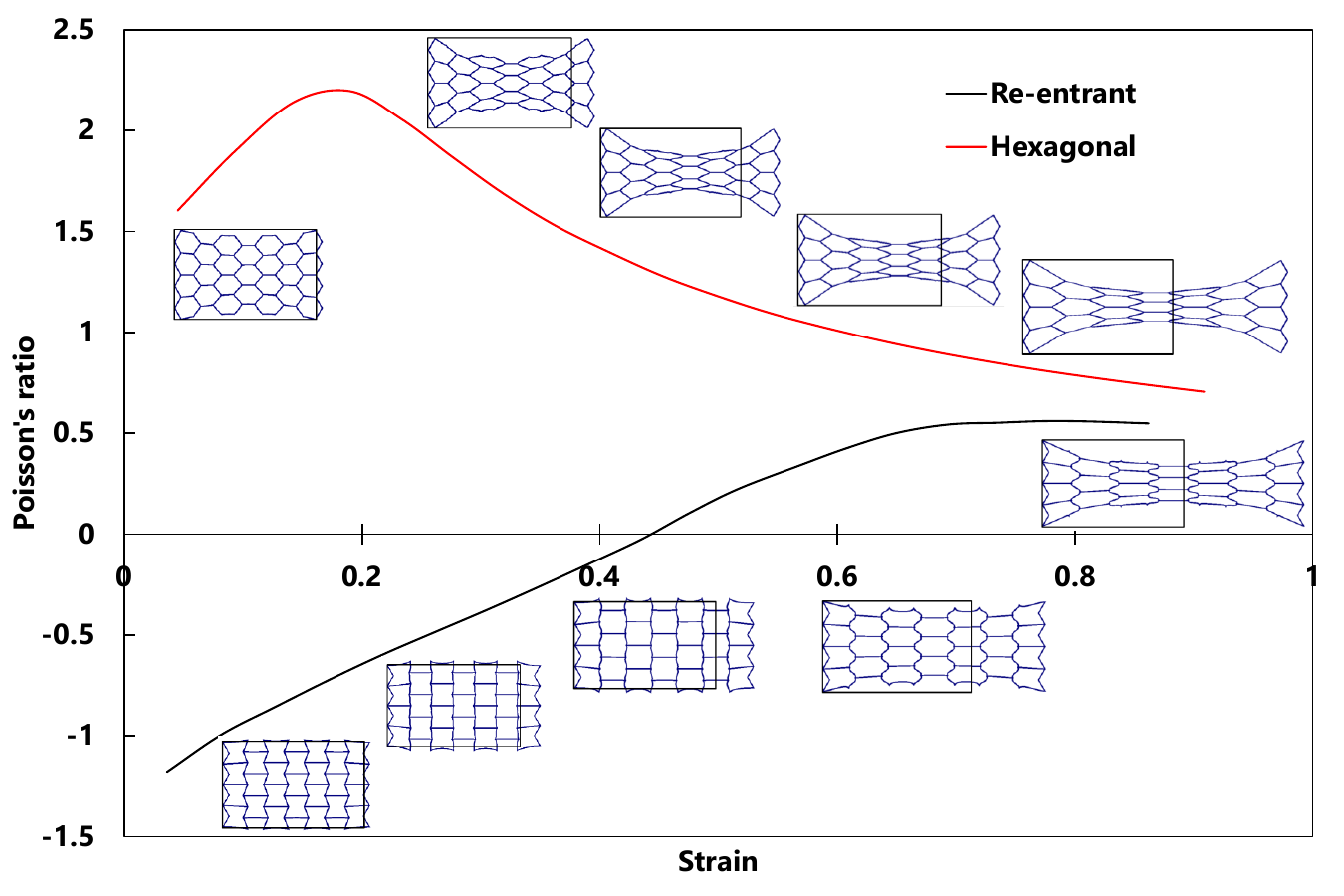}
                        \vspace{-0.5cm}
                        \caption{}
                        \label{fig12b}
                    \vspace{-0.1cm}
                \end{subfigure}
                \vspace{0.0cm}
                \caption{Poisson's ratios of RA and HH structures under (a) y and (b) x directional tension.}
                \label{fig12}
            \end{figure}

            \autoref{fig12a} and \autoref{fig12b} display the Poisson's ratio-strain curve and deformation patterns for tension in the y and x directions, respectively. The change of Poisson's ratio is different when subjected to tension. when the structure is stretched the rate of Poisson's ratio is greater compared to compression. In the hexagonal lattice structure, the Poisson's ratio is greater than one when strain is less than 0.2 (\autoref{fig12}). This goes against what we usually see in regular materials, where Poisson's ratio typically ranges from zero to 0.5 \cite{hao2023novel}. However, lattice structures can exhibit tunable Poisson's ratios, which can be either less than zero or greater than one \cite{montgomery2023elastic}. As shown in \autoref{fig12b}, when the re-entrant structure is stretched, it expands laterally up to the strain $\varepsilon = 0.44$; thereafter, it contracts. The transition in the Poisson's ratio from negative to positive values causes the re-entrant structure to exhibit behavior similar to conventional materials after significant stretching under tensile loading. The ability to create a wide range of Poisson's ratios enables these structures to be applied in various impact resistance applications \cite{bohara2023anti}.

%------------------------------------------------------------------------------

	\subsection{Energy absorption}
	\label{subsec3.3}

            Energy absorption is one of the remarkable mechanical characteristics of the auxetic structures due to its capability to reduce the impact forces in dynamic loading. Auxetic structures are known to absorb and disperse energy more than traditional structures. This section provides an insight into the analysis of the energy absorption capability of such structures. The total amount of the energy absorption is the integration of the force-displacement curve and can be expressed as:
    
            \begin{equation}\label{eq22}
                {{E}_{a}}=\int_{0}^{l}{Pdl}
            \end{equation}
    where $l$ is the displacement corresponding to the densification strain and $P$ is the force at displacement $l$. The densification strain is a crucial factor in assessing the energy absorption capabilities of lattice structures. It is related to the state that the cell walls are folded entirely and accumulated over each other. After the densification strain, the structure no longer bears the applied load and transfers it. Therefore, the energy absorption is computed up to the densification strain point, identified as the strain value at the peak of the energy efficiency curve. The energy efficiency can be calculated as:
    
            \begin{equation}\label{eq23}
                {{E}_{F}}({{\varepsilon }_{a}})=\frac{\int_{0}^{{{\varepsilon }_{a}}}{\sigma (\varepsilon )d\varepsilon }}{{{\sigma }_{a}}}\,,\,\,\,\,\,\,\,\,\,0\le {{\varepsilon }_{a}}<1
            \end{equation}
    where ${{\varepsilon }_{a}}$ is the strain at the certain moment and ${{\sigma }_{a}}$ is the stress at ${{\varepsilon }_{a}}$. The densification strain (${{\varepsilon }_{a}}$) is calculated using the following equation:
    
            \begin{equation}\label{eq24}
                \frac{d{{E}_{f}}({{\varepsilon }_{a}})}{d{{\varepsilon }_{a}}}{{|}_{{{\varepsilon }_{a}}={{\varepsilon }_{d}}}}=0\,\,\,\,\,\,0\le {{\varepsilon }_{a}}<1
            \end{equation}
    
\begin{figure}[h]%
                \centering
                \includegraphics[width=0.6\textwidth]{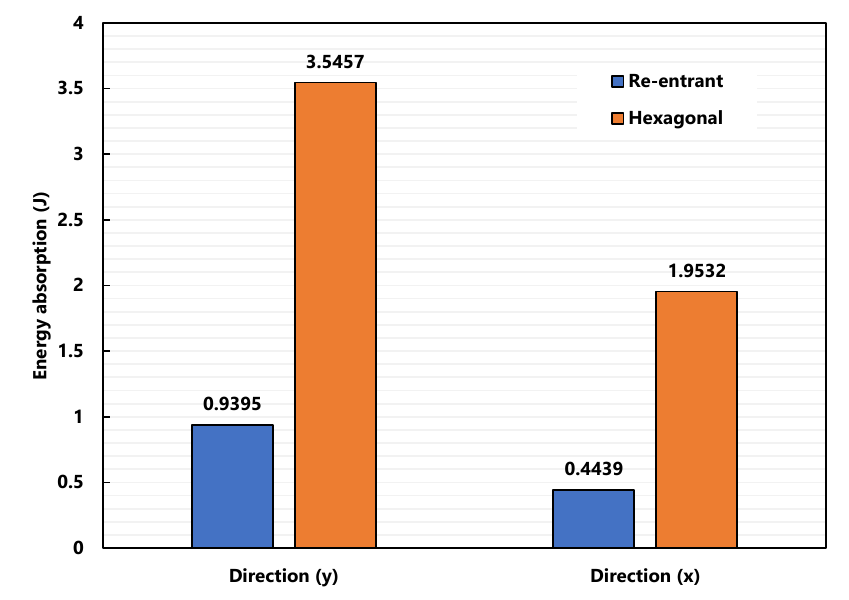}
                \caption{Energy absorption.}
                \label{fig13}  
            \end{figure}

\begin{figure}[!]%
                \centering
                \includegraphics[width=0.6\textwidth]{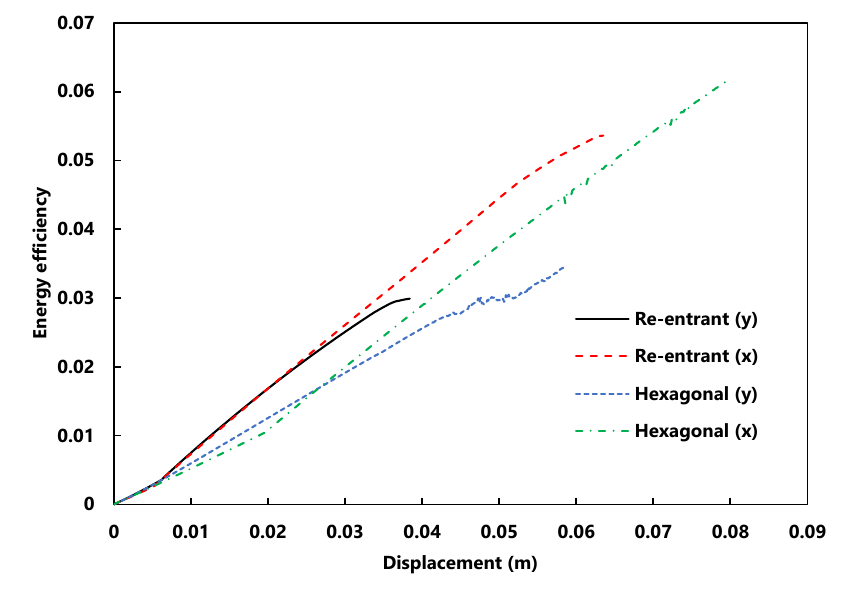}
                \caption{Efficiency of the energy absorption.}
                \label{fig14}  
            \end{figure}

            To assess the energy absorption capacity of the structures, several factors are considered, these include the area under the force-displacement curve, the energy absorption efficiency, and the densification strain. It should be noted that the densification strains highlighted in \autoref{fig7b} and \autoref{fig8b} are calculated using \autoref{eq24}. From \autoref{fig7b} and \autoref{fig8b}, it's evident that the area under the stress-strain curve for the honeycomb structure is greater than that for the re-entrant structure. This implies that the honeycomb structure exhibits higher energy absorption compared to the re-entrant structure \cite{luo2022mechanical}. This is demonstrated in \autoref{fig13} where the energy absorption for the structures under compression is presented. The energy absorption is higher for the honeycomb structure, particularly in the case of y-axis compression. The re-entrant structure absorbs more energy in y-directional compression compared to x-directional loading. However, it should be noted that during the plateau stage of the re-entrant structures, constant stress may affect positively the energy absorption capacity. For this reason, the energy absorption efficiency of the structures is calculated and presented in \autoref{fig14}. It can be observed that the energy absorption efficiency of the re-entrant structure is greater than the honeycomb structure at the same displacement. This implies that re-entrant structures can absorb energy over an extended duration, without an immediate response to applied force. By having this ability to absorb energy gradually, re-entrant structures can prevent a rapid increase in applied force. This characteristic is beneficial in situations where sudden force acts on a structure. Instead of an immediate and potentially destructive response, the re-entrant design allows for a more controlled and gradual absorption of energy, helping to dissipate and distribute the forces more effectively. This characteristic makes them suitable for protective applications.

 %------------------------------------------------------------------------------
 %------------------------------------------------------------------------------

    \section{Conclusions}
    \label{sec4}

        This study investigates the mechanical characteristics of HH and RA structures under compression and tension in both the y and x directions. 2D plane strain and 3D nonlinear large deformation behaviors are analyzed using the hyperelastic-plastic constitutive model. The model parameters are calibrated based on experimental data. The auxetic and non-auxetic structures are explored and their structural behaviour under different loading conditions are assessed. Based on the obtained results, several conclusions can be drawn as below:

        (1) Under in-plane loading, the 3D analysis of the structures can be simplified with minimal error using a 2D plane strain assumption. The similarity in stress response and deformation patterns between 2D and 3D numerical results supports this simplification. 

        (2) The deformation patterns of the RA and HH structures differ. The compression stress-strain curve of the re-entrant structure is divided into three parts: elastic, plateau, and densification. In the plateau stage, the stress stays relatively constant as the strain increases crucial in preventing a rapid increase in impact force. In contrast, honeycomb structures display a steady rise in stress as strain increases.

        (3) RA structures exhibit a negative Poisson's ratio under both compression and tension approaching zero as strain increases. In contrast, HH structures show an opposite behavior, with the Poisson's ratio decreasing from one to zero.

        (4) The energy absorption efficiency is higher in RA structures, indicating their ability to absorb energy over an extended duration and mitigate the rapid increase in applied force.

        (5) Both structures demonstrate increased stiffness when subjected to y-directional loading, allowing them to tolerate higher stresses. Notably, the auxetic structure loses its characteristic behavior as strain increases during tensile loading, highlighting the significant influence of applied force direction on the mechanical behavior of these structures.

 %------------------------------------------------------------------------------
 %------------------------------------------------------------------------------

    \section*{Declaration of competing interest}
	
	The authors declare that they have no known competing financial interests or personal relationships that could have appeared to influence the work reported in this paper.

 %------------------------------------------------------------------------------
 %------------------------------------------------------------------------------

    \section*{Acknowledgments}
	
	The authors acknowledge the support received by the PriMus project with the grant PID2020-  115575RB-I00  funded by MCIN/AEI/ 10.13039/501100011033/.

	\bibliography{Manuscript}

\begin{thebibliography}{48}
\providecommand{\natexlab}[1]{#1}
\providecommand{\url}[1]{\texttt{#1}}
\expandafter\ifx\csname urlstyle\endcsname\relax
  \providecommand{\doi}[1]{doi: #1}\else
  \providecommand{\doi}{doi: \begingroup \urlstyle{rm}\Url}\fi

\bibitem[Rivet et~al.(2021)Rivet, Dialami, Cervera, Chiumenti, Reyes, and
  P{\'e}rez]{rivet2021experimental}
Iv{\'a}n Rivet, Narges Dialami, Miguel Cervera, Michele Chiumenti, Guillermo
  Reyes, and Marco~A P{\'e}rez.
\newblock Experimental, computational, and dimensional analysis of the
  mechanical performance of fused filament fabrication parts.
\newblock \emph{Polymers}, 13\penalty0 (11):\penalty0 1766, 2021.

\bibitem[Dialami et~al.(2022)Dialami, Chiumenti, Cervera, Chasco, Reyes-Pozo,
  and P{\'e}rez]{dialami2022hybrid}
Narges Dialami, Michele Chiumenti, Miguel Cervera, Uxue Chasco, Guillermo
  Reyes-Pozo, and Marco~A P{\'e}rez.
\newblock A hybrid numerical-experimental strategy for predicting mechanical
  response of components manufactured via fff.
\newblock \emph{Composite Structures}, 298:\penalty0 115998, 2022.

\bibitem[Dialami et~al.(2021)Dialami, Chiumenti, Cervera, Rossi, Chasco, and
  Domingo]{dialami2021numerical}
Narges Dialami, Michele Chiumenti, Miguel Cervera, Riccardo Rossi, Uxue Chasco,
  and Miquel Domingo.
\newblock Numerical and experimental analysis of the structural performance of
  am components built by fused filament fabrication.
\newblock \emph{International journal of mechanics and materials in design},
  17:\penalty0 225--244, 2021.

\bibitem[Wu et~al.(2023)Wu, Gao, and Xiong]{wu2023quasi}
Qianqian Wu, Ying Gao, and Jian Xiong.
\newblock Quasi-static mechanical properties of composite lattice sandwich
  structures with enhanced face panels.
\newblock \emph{European Journal of Mechanics-A/Solids}, 97:\penalty0 104808,
  2023.

\bibitem[Dialami et~al.(2023)Dialami, Rivet, Cervera, and
  Chiumenti]{dialami2023computational}
Narges Dialami, Ivan Rivet, Miguel Cervera, and Michele Chiumenti.
\newblock Computational characterization of polymeric materials 3d-printed via
  fused filament fabrication.
\newblock \emph{Mechanics of Advanced Materials and Structures}, 30\penalty0
  (7):\penalty0 1357--1367, 2023.

\bibitem[Shirzad et~al.(2022)Shirzad, Zolfagharian, Bodaghi, and
  Nam]{shirzad2022auxetic}
Masoud Shirzad, Ali Zolfagharian, Mahdi Bodaghi, and Seung~Yun Nam.
\newblock Auxetic metamaterials for bone-implanted medical devices: recent
  advances and new perspectives.
\newblock \emph{European Journal of Mechanics-A/Solids}, page 104905, 2022.

\bibitem[Ingrole et~al.(2017)Ingrole, Hao, and Liang]{ingrole2017design}
Aniket Ingrole, Ayou Hao, and Richard Liang.
\newblock Design and modeling of auxetic and hybrid honeycomb structures for
  in-plane property enhancement.
\newblock \emph{Materials \& Design}, 117:\penalty0 72--83, 2017.

\bibitem[Xu et~al.(2020)Xu, Liu, Wang, Zhang, Jia, Lei, and Fang]{xu2020plane}
Mengchuan Xu, Debo Liu, Panding Wang, Zhong Zhang, Heran Jia, Hongshuai Lei,
  and Daining Fang.
\newblock In-plane compression behavior of hybrid honeycomb metastructures:
  Theoretical and experimental studies.
\newblock \emph{Aerospace Science and Technology}, 106:\penalty0 106081, 2020.

\bibitem[Evans and Alderson(2000)]{evans2000auxetic}
Kenneth~E Evans and KL~Alderson.
\newblock Auxetic materials: the positive side of being negative.
\newblock \emph{Engineering Science \& Education Journal}, 9\penalty0
  (4):\penalty0 148--154, 2000.

\bibitem[Zhou et~al.(2024)Zhou, Gao, Wang, Zheng, Lv, Ma, Sun, and
  Wang]{zhou2024energy}
Jianzhong Zhou, Qiang Gao, Liangmo Wang, Xuyang Zheng, Hao Lv, Zhiyong Ma,
  Huiming Sun, and Xiaoyu Wang.
\newblock Energy absorption of a novel auxetic structure reinforced by
  embedding tubes.
\newblock \emph{European Journal of Mechanics-A/Solids}, page 105338, 2024.

\bibitem[Krishnan et~al.(2022)Krishnan, Biswas, Kumar, and
  Sreekanth]{krishnan2022auxetic}
Bharath~R Krishnan, Ankan~Narayan Biswas, KV~Ahalya Kumar, and PS~Rama
  Sreekanth.
\newblock Auxetic structure metamaterial for crash safety of sports helmet.
\newblock \emph{Materials Today: Proceedings}, 56:\penalty0 1043--1049, 2022.

\bibitem[Wang and Hu(2014)]{wang2014auxetic}
Zhengyue Wang and Hong Hu.
\newblock Auxetic materials and their potential applications in textiles.
\newblock \emph{Textile Research Journal}, 84\penalty0 (15):\penalty0
  1600--1611, 2014.

\bibitem[Alomarah et~al.(2023)Alomarah, Yuan, and Ruan]{alomarah2023bio}
Amer Alomarah, Ye~Yuan, and Dong Ruan.
\newblock A bio-inspired auxetic metamaterial with two plateau regimes:
  Compressive properties and energy absorption.
\newblock \emph{Thin-Walled Structures}, 192:\penalty0 111175, 2023.

\bibitem[Chan and Evans(1999)]{chan1999mechanical}
N~Chan and KE~Evans.
\newblock The mechanical properties of conventional and auxetic foams. part i:
  compression and tension.
\newblock \emph{Journal of Cellular plastics}, 35\penalty0 (2):\penalty0
  130--165, 1999.

\bibitem[Choi and Lakes(1992)]{choi1992non}
JB~Choi and RS~Lakes.
\newblock Non-linear properties of metallic cellular materials with a negative
  poisson's ratio.
\newblock \emph{Journal of Materials Science}, 27:\penalty0 5375--5381, 1992.

\bibitem[Friis et~al.(1988)Friis, Lakes, and Park]{friis1988negative}
EA~Friis, RS~Lakes, and JB~Park.
\newblock Negative poisson's ratio polymeric and metallic foams.
\newblock \emph{Journal of Materials science}, 23:\penalty0 4406--4414, 1988.

\bibitem[Galati et~al.(2022)Galati, Calignano, and Minosi]{galati2022numerical}
Manuela Galati, Flaviana Calignano, and Francesco Minosi.
\newblock Numerical and experimental investigations of a novel 3d bucklicrystal
  auxetic structure produced by metal additive manufacturing.
\newblock \emph{Thin-Walled Structures}, 180:\penalty0 109850, 2022.

\bibitem[Dominec et~al.(1992)Dominec, Va{\v{s}}ek, Svoboda,
  Plech{\'a}{\v{c}}ek, and Laermans]{dominec1992elastic}
J~Dominec, P~Va{\v{s}}ek, P~Svoboda, V~Plech{\'a}{\v{c}}ek, and C~Laermans.
\newblock Elastic moduli for three superconducting phases of bi-sr-ca-cu-o.
\newblock \emph{Modern Physics Letters B}, 6\penalty0 (16n17):\penalty0
  1049--1054, 1992.

\bibitem[Zhang et~al.(2015)Zhang, Ghita, and Evans]{zhang2015fabrication}
GH~Zhang, O~Ghita, and Kenneth~E Evans.
\newblock The fabrication and mechanical properties of a novel 3-component
  auxetic structure for composites.
\newblock \emph{Composites Science and Technology}, 117:\penalty0 257--267,
  2015.

\bibitem[Li et~al.(2023{\natexlab{a}})Li, Wang, Ma, Wu, and
  Wang]{li2023auxetic}
Zhen-Yu Li, Xin-Tao Wang, Li~Ma, Lin-Zhi Wu, and Lifeng Wang.
\newblock Auxetic and failure characteristics of composite stacked origami
  cellular materials under compression.
\newblock \emph{Thin-Walled Structures}, 184:\penalty0 110453,
  2023{\natexlab{a}}.

\bibitem[Zhou et~al.(2015)Zhou, Deng, Yan, Chen, and
  Liu]{zhou2015superelasticity}
Jialv Zhou, Xiaobin Deng, Yuan Yan, Xi~Chen, and Yilun Liu.
\newblock Superelasticity and reversible energy absorption of polyurethane
  cellular structures with sand filler.
\newblock \emph{Composite Structures}, 131:\penalty0 966--974, 2015.

\bibitem[Abel et~al.(2021)Abel, Mannschatz, Teuber, M{\"u}ller, Al~Noaimy,
  Riecker, Thielsch, Matthey, and Wei{\ss}g{\"a}rber]{abel2021fused}
Johannes Abel, Anne Mannschatz, Robert Teuber, Bernhard M{\"u}ller, Omar
  Al~Noaimy, Sebastian Riecker, Juliane Thielsch, Bj{\"o}rn Matthey, and Thomas
  Wei{\ss}g{\"a}rber.
\newblock Fused filament fabrication of niti components and hybridization with
  laser powder bed fusion for filigree structures.
\newblock \emph{Materials}, 14\penalty0 (16):\penalty0 4399, 2021.

\bibitem[Hu and Wang(2021)]{hu2021crack}
JS~Hu and BL~Wang.
\newblock Crack growth behavior and thermal shock resistance of ceramic
  sandwich structures with an auxetic honeycomb core.
\newblock \emph{Composite Structures}, 260:\penalty0 113256, 2021.

\bibitem[Yu et~al.(2023)Yu, Liu, Cao, Liu, Huang, and Wang]{yu2023compressive}
Sheng Yu, Zhikang Liu, Xiaoming Cao, Jiayi Liu, Wei Huang, and Yangwei Wang.
\newblock The compressive responses and failure behaviors of composite graded
  auxetic re-entrant honeycomb structure.
\newblock \emph{Thin-Walled Structures}, 187:\penalty0 110721, 2023.

\bibitem[Gibson and Ashby(1982)]{gibson1982mechanics}
IJ~Gibson and Michael~Farries Ashby.
\newblock The mechanics of three-dimensional cellular materials.
\newblock \emph{Proceedings of the royal society of London. A. Mathematical and
  physical sciences}, 382\penalty0 (1782):\penalty0 43--59, 1982.

\bibitem[Larsen et~al.(1997)Larsen, Signund, and Bouwsta]{larsen1997design}
Ulrik~Darling Larsen, O~Signund, and S~Bouwsta.
\newblock Design and fabrication of compliant micromechanisms and structures
  with negative poisson's ratio.
\newblock \emph{Journal of microelectromechanical systems}, 6\penalty0
  (2):\penalty0 99--106, 1997.

\bibitem[Theocaris et~al.(1997)Theocaris, Stavroulakis, and
  Panagiotopoulos]{theocaris1997negative}
PS~Theocaris, GE~Stavroulakis, and PD~Panagiotopoulos.
\newblock Negative poisson's ratios in composites with star-shaped inclusions:
  a numerical homogenization approach.
\newblock \emph{Archive of applied mechanics}, 67:\penalty0 274--286, 1997.

\bibitem[Montgomery-Liljeroth et~al.(2023)Montgomery-Liljeroth, Schievano, and
  Burriesci]{montgomery2023elastic}
Ebba Montgomery-Liljeroth, Silvia Schievano, and Gaetano Burriesci.
\newblock Elastic properties of 2d auxetic honeycomb structures-a review.
\newblock \emph{Applied Materials Today}, 30:\penalty0 101722, 2023.

\bibitem[Lakes(1991)]{lakes1991deformation}
Roderic Lakes.
\newblock Deformation mechanisms in negative poisson's ratio materials:
  structural aspects.
\newblock \emph{Journal of materials science}, 26:\penalty0 2287--2292, 1991.

\bibitem[Li et~al.(2023{\natexlab{b}})Li, Liu, Wu, Wang, Han, and
  Zhang]{li2023mechanical}
Na~Li, Shu-zun Liu, Xiao-nan Wu, Jun-yu Wang, Yue-song Han, and Xin-chun Zhang.
\newblock Mechanical characteristics of a novel rotating star-rhombic auxetic
  structure with multi-plateau stages.
\newblock \emph{Thin-Walled Structures}, 191:\penalty0 111081,
  2023{\natexlab{b}}.

\bibitem[Teng et~al.(2023)Teng, Jiang, Zhang, Han, Ni, Xu, Hao, Guo, Wu, Xie,
  et~al.]{teng2023stretchable}
Xing~Chi Teng, Wei Jiang, Xue~Gang Zhang, Dong Han, Xi~Hai Ni, Hang~Hang Xu,
  Jian Hao, Tong Guo, Yu~Fei Wu, Yi~Min Xie, et~al.
\newblock A stretchable sandwich panel metamaterial with auxetic
  rotating-square surface.
\newblock \emph{International Journal of Mechanical Sciences}, 251:\penalty0
  108334, 2023.

\bibitem[Tao et~al.(2022)Tao, Ren, Zhao, Sun, Zhang, Jiang, Han, Zhang, and
  Xie]{tao2022novel}
Zhi Tao, Xin Ren, Ai~Guo Zhao, Long Sun, Yi~Zhang, Wei Jiang, Dong Han,
  Xiang~Yu Zhang, and Yi~Min Xie.
\newblock A novel auxetic acoustic metamaterial plate with tunable bandgap.
\newblock \emph{International Journal of Mechanical Sciences}, 226:\penalty0
  107414, 2022.

\bibitem[Grima and Evans(2000)]{grima2000auxetic}
Joseph~N Grima and Kenneth~E Evans.
\newblock Auxetic behavior from rotating squares.
\newblock \emph{Journal of materials science letters}, 19:\penalty0 1563--1565,
  2000.

\bibitem[Zhou et~al.(2023)Zhou, Liu, Dear, Falzon, and
  Kazanc{\i}]{zhou2023comparison}
Jin Zhou, Haibao Liu, John~P Dear, Brian~G Falzon, and Zafer Kazanc{\i}.
\newblock Comparison of different quasi-static loading conditions of additively
  manufactured composite hexagonal and auxetic cellular structures.
\newblock \emph{International Journal of Mechanical Sciences}, 244:\penalty0
  108054, 2023.

\bibitem[Luo et~al.(2022)Luo, Ren, Zhang, Zhang, Zhang, Luo, Cheng, and
  Xie]{luo2022mechanical}
Hui~Chen Luo, Xin Ren, Yi~Zhang, Xiang~Yu Zhang, Xue~Gang Zhang, Chen Luo, Xian
  Cheng, and Yi~Min Xie.
\newblock Mechanical properties of foam-filled hexagonal and re-entrant
  honeycombs under uniaxial compression.
\newblock \emph{Composite Structures}, 280:\penalty0 114922, 2022.

\bibitem[Carbonell et~al.(2020)Carbonell, Rodr{\'\i}guez, and
  O{\~n}ate]{carbonell2020modelling}
Josep~Maria Carbonell, JM~Rodr{\'\i}guez, and Eugenio O{\~n}ate.
\newblock Modelling 3d metal cutting problems with the particle finite element
  method.
\newblock \emph{Computational Mechanics}, 66:\penalty0 603--624, 2020.

\bibitem[Dohrmann and Bochev(2004)]{dohrmann2004stabilized}
Clark~R Dohrmann and Pavel~B Bochev.
\newblock A stabilized finite element method for the stokes problem based on
  polynomial pressure projections.
\newblock \emph{International Journal for Numerical Methods in Fluids},
  46\penalty0 (2):\penalty0 183--201, 2004.

\bibitem[Rodriguez~Prieto et~al.(2018)Rodriguez~Prieto, Carbonell, Cante,
  Oliver, and Jons{\'e}n]{rodriguez2018generation}
Juan~Manuel Rodriguez~Prieto, Josep~Maria Carbonell, JC~Cante, Javier Oliver,
  and P{\"a}r Jons{\'e}n.
\newblock Generation of segmental chips in metal cutting modeled with the pfem.
\newblock \emph{Computational Mechanics}, 61:\penalty0 639--655, 2018.

\bibitem[Rodriguez et~al.(2016)Rodriguez, Carbonell, Cante, and
  Oliver]{rodriguez2016particle}
JM~Rodriguez, Josep~Maria Carbonell, JC~Cante, and J35402641352 Oliver.
\newblock The particle finite element method (pfem) in thermo-mechanical
  problems.
\newblock \emph{International journal for numerical methods in engineering},
  107\penalty0 (9):\penalty0 733--785, 2016.

\bibitem[Rodr{\'\i}guez et~al.(2017)Rodr{\'\i}guez, Carbonell, Cante, and
  Oliver]{rodriguez2017continuous}
Juan~Manuel Rodr{\'\i}guez, Josep~Maria Carbonell, JC~Cante, and Javier Oliver.
\newblock Continuous chip formation in metal cutting processes using the
  particle finite element method (pfem).
\newblock \emph{International Journal of Solids and Structures}, 120:\penalty0
  81--102, 2017.

\bibitem[Simo and Miehe(1992)]{simo1992associative}
JC~Simo and Ch~Miehe.
\newblock Associative coupled thermoplasticity at finite strains: Formulation,
  numerical analysis and implementation.
\newblock \emph{Computer Methods in Applied Mechanics and Engineering},
  98\penalty0 (1):\penalty0 41--104, 1992.

\bibitem[de~Souza~Neto et~al.(2011)de~Souza~Neto, Peric, and
  Owen]{de2011computational}
Eduardo~A de~Souza~Neto, Djordje Peric, and David~RJ Owen.
\newblock \emph{Computational methods for plasticity: theory and applications}.
\newblock John Wiley \& Sons, 2011.

\bibitem[Oliver et~al.(2009)Oliver, Hartmann, Cante, Weyler, and
  Hernandez]{oliver2009contact}
Javier Oliver, S~Hartmann, JC~Cante, R~Weyler, and JA2536080 Hernandez.
\newblock A contact domain method for large deformation frictional contact
  problems. part 1: Theoretical basis.
\newblock \emph{Computer methods in applied mechanics and engineering},
  198\penalty0 (33-36):\penalty0 2591--2606, 2009.

\bibitem[Hartmann et~al.(2009)Hartmann, Oliver, Weyler, Cante, and
  Hern{\'a}ndez]{hartmann2009contact}
S~Hartmann, Javier Oliver, R~Weyler, JC~Cante, and JA~Hern{\'a}ndez.
\newblock A contact domain method for large deformation frictional contact
  problems. part 2: Numerical aspects.
\newblock \emph{Computer Methods in Applied Mechanics and Engineering},
  198\penalty0 (33-36):\penalty0 2607--2631, 2009.

\bibitem[Melendo et~al.(2018)Melendo, Coll, Pasenau, Escolano, and
  Monros]{gidhome}
A.~Melendo, A.~Coll, M.~Pasenau, E.~Escolano, and A.~Monros.
\newblock www.gidhome.com, 2018.
\newblock
  \href{https://www.gidsimulation.com/}{https://www.gidsimulation.com/}.

\bibitem[Carbonell()]{PFEM}
JM~Carbonell.
\newblock Pfem-kratos: Development for the kinetics of fluids, solids and
  grounds, modelling of subtractive manufacturing processes.
\newblock \emph{GitLab repository:
  \href{https://gitlab.com/pfem-research/kratos}{https://gitlab.com/pfem-research/kratos}}.

\bibitem[Hao et~al.(2023)Hao, Han, Zhang, Teng, Xu, Jiang, Lang, Ni, Luo, Li,
  et~al.]{hao2023novel}
Jian Hao, Dong Han, Xue~Gang Zhang, Xing~Chi Teng, Hang~Hang Xu, Wei Jiang,
  Jian~Ping Lang, Xi~Hai Ni, Yu~Ming Luo, Hao~Ran Li, et~al.
\newblock A novel compression-torsion coupling metamaterial with tunable
  poisson's ratio.
\newblock \emph{Construction and Building Materials}, 395:\penalty0 132276,
  2023.

\bibitem[Bohara et~al.(2023)Bohara, Linforth, Nguyen, Ghazlan, and
  Ngo]{bohara2023anti}
Rajendra~Prasad Bohara, Steven Linforth, Tuan Nguyen, Abdallah Ghazlan, and
  Tuan Ngo.
\newblock Anti-blast and-impact performances of auxetic structures: A review of
  structures, materials, methods, and fabrications.
\newblock \emph{Engineering Structures}, 276:\penalty0 115377, 2023.

\end{thebibliography}
	
\end{document}